\documentclass[printer]{aa} 
\usepackage{graphicx}
\usepackage{natbib}
\usepackage{amsmath}
\usepackage{amscd}
\usepackage{booktabs}
\usepackage{caption}
\usepackage{multirow}
\usepackage{subfigure}
\usepackage{url}
\usepackage{subfigure}
\usepackage{bm}
\usepackage{color}
\usepackage{txfonts}
\usepackage{lineno}
\bibpunct{(}{)}{;}{a}{}{,} 
\newcommand{\Gaia}{\emph{Gaia}}
\newcommand{\teff}{$T_{\rm eff}$}
\newcommand{\logg}{$\log(g)$}
\begin{document}
   \title{Properties of ultra-cool dwarfs with {\Gaia}}
   \subtitle{An assessment of the accuracy for the temperature determination}
   \author{L.M. Sarro
          \inst{1}
          \and
	  A. Berihuete\inst{2}
	  \and
	  C. Carri\'on\inst{1}
	  \and
          D. Barrado\inst{3,4}
	  \and
	  P. Cruz\inst{3}
	  \and
	  Y. Isasi\inst{5}
          }

   \institute{Dpt. de Inteligencia Artificial , UNED, Juan del Rosal,
     16, 28040 Madrid, Spain\\ \email{lsb@dia.uned.es} 
     \and
     Dpt. Statistics and Operations Research, University of
     C\'adiz, Campus Universitario R\'io San Pedro s/n.  11510 Puerto
     Real, C\'adiz, Spain\\ \email{angel.berihuete@uca.es}
     \and
     Calar Alto Observatory, Centro Astron\'omico Hispano Alem\'an, C/ Jes\'us Durb\'an Rem\'on,
     E-04004 Almer\'{\i}a, Spain 
     \and
     Depto. Astrof\'{\i}sica, Centro de Astrobiolog\'{\i}a (INTA-CSIC), ESAC campus,   P.O. Box 78,
     E-28691 Villanueva de la Ca\~nada, Spain 
     \and
     Department Astronomia i Meteorologia ICCUB-IEEC, Mart\'{\i} i Franqu\`es 1, Barcelona, 08028 Spain 
   }

   \date{}
 
  \abstract
  {The {\Gaia} catalogue will contain observations and physical
    parameters of a vast number of objects, including ultra-cool dwarf
    stars, which we define here as stars with a temperature below 2500 K.}
   {We aimed to assess the accuracy of the Gaia {\teff} and {\logg}
     estimates as derived with current models and observations.}
{We assessed the validity of several inference techniques for deriving the
  physical parameters of ultra-cool dwarf stars: Gaussian processes,
  support vector machines, $k$-nearest neighbours, kernel partial
  least squares and Bayesian estimation. In addition, we tested the
  potential benefits of data compression for improving robustness and
  speed. We used synthetic spectra derived from ultra-cool dwarf
    models to construct (train) the regression models. We derived the
  intrinsic uncertainties of the best inference models and assessed
  their validity by comparing the estimated parameters with the values
  derived in the bibliography for a sample of ultra-cool dwarf stars
  observed from the ground.}
{{\bf We estimated the total number of ultra-cool dwarfs per spectral
    subtype, and obtained values that can be summarised (in orders of
    magnitude) as 400000 objects in the M5-L0 range, 600 objects
    between L0 and L5, 30 objects between L5 and T0, and 10 objects
    between T0 and T8. A bright ultra-cool dwarf (with {\teff}=2500 K
    and {\logg}=3.5) will be detected by {\Gaia} out to approximately
    220 pc, while for {\teff}=1500 K (spectral type L5) and the same
    surface gravity, this maximum distance reduces to 10-20 pc.}  We
  found the cross-validation RMSE prediction error to be 10 K for
  regression models based on the $k$-nearest neighbours and 62 K for
  Gaussian process models in the faintest limit ({\Gaia} magnitude
  G=20). However these values correspond to the evaluation
  of the regression models with independent test sets of synthetic
  spectra of the same model families as used in the training phase
  (internal errors).  For the $k$-nearest neighbours model,
  this seems an overly optimistic error estimate due to the use of a
  dense grid of examples in the training set, together with a
  relatively high signal-to-noise ratio for the end-of-mission
  data. The RMSE of the prediction deduced from ground-based spectra
  of ultra-cool dwarfs simulated at the {\Gaia} spectral range and
  resolution, and for a {\Gaia} magnitude G=20 is 213 K and 266 K for
  the models based on $k$-nearest neighbours and Gaussian process
  regression, respectively. These are total errors in the sense that they
  include the internal and external errors, with the latter caused by the
  inability of the synthetic spectral models (used for the
  construction of the regression models) to exactly reproduce the
  observed spectra, and by the large uncertainties in the current
  calibrations of spectral types and effective temperatures.  We found
  maximum-likelihood methods (minimum $\chi^2$, $k$-nearest neighbours,
  and Bayesian estimation with flat priors) to be biased in the L0-T0
  range in that they systematically assign a temperature around
  1700 K. Finally, the likelihood landscape is significantly
  multimodal in spectra with realistic noise.}
   {}
   \keywords{Methods: data analysis,Methods: statistical, Catalogues,
     brown dwarfs, Stars: fundamental parameters}
   \maketitle
%


\section{Introduction}
\label{intro}

In this work we define an ultra-cool star as a star with an effective
temperature below 2500 K (spectral type M8). The goal of this paper is
to assess the detectability of this type of object with the Gaia
spacecraft.  Gaia is a mission of the European Space Agency that will
produce very accurate astrometry and parallaxes for a significant
fraction of the galactic population \citep{db2012}, thus helping to
considerably improve our knowledge of a plethora of astronomical
topics, from stellar evolution to exoplanets.  In particular, {\Gaia}
data will improve our understanding of the nature of ultra-cool dwarfs
by providing distances and therefore luminosities for the nearest
objects.  The 2MASS All-Sky survey (Cutri et al. 1996) began a new era
in the discovery and characterisation of very-late spectral type stars
and brown dwarfs, allowing the identification of two new types: L
(Kirpatrick et al. 1999; Mart\'{\i}n et al. 1997) and T (Kirpatrick et
al. 1999; Burgasser et al. 2002), and paving the way for even cooler
objects, the Y type (Burningham et al. 2008; Kirkpatrick et
al. 2012). Subsequent surveys have discovered hundreds of cool
objects, but the comprehensive understanding of their nature relies in
modelling from internal structure to atmospheres. This can only be
achieved with precise data, including accurate distances. Here, Gaia
will truly play the role of a Rosetta stone.

The work presented here was developed in the framework of the eighth
coordination unit (CU8) of the {\Gaia} Data Processing and
Analysis Consortium
(DPAC\footnote{\url{http://www.rssd.esa.int/index.php?project=GAIA&page=DPAC_Introduction}}), which is devoted to determining astrophysical parameters.  {\Gaia} is
expected to detect and characterise one billion sources, and hence,
automatic procedures for the reduction and processing of these data
are essential. 

The DPAC consortium is, in charge of the design, development, and
operation of this data processing and analysis chain aimed of
producing the {\Gaia} catalogue (intermediate releases and final
catalogue) from the telemetry data (see \cite{DPAC} for a more detailed
introduction to the DPAC).

Since potential {\Gaia} targets include very different
astrophysical scenarios, from unresolved galaxies and quasars to
asteroids, specialised modules have been designed and implemented
within the DPAC to characterise the various object types.

The astrophysical parameters in CU8 are determined by
various modules integrated in the the {\sl Apsis} pipeline. {\sl
  Apsis} includes an initial classification into broad object
categories \citep{DSC}, modules for derivating stellar
parameters from (amongst other observables) very low- and
medium-resolution spectra \citep[][respectively]{GSPPHOT,RVS}, and
specialised modules for characterising of unresolved galaxies
\citep{galaxies}, quasars, or peculiar types of stars such as emission-line stars \citep{ELS}, or cool stars.

In particular, a specific module of the Gaia processing pipeline is
devoted to characterising ultra-cool dwarfs (hereafter UCDs),
which constitutes a regression problem in which the source parameters
are estimated from observational data.  In the next section, we describe this module and estimate the number
of sources that will be detected as a function of spectral type. In Sect.  \ref{method} we briefly describe the statistical
techniques explored in the search for an optimum model for the
regression problem of determining the source parameters, together with
the experiments carried out in order to select amongst them. In
Sect. \ref{valid} we describe the results obtained by these models
when they are applied to simulated Gaia spectra of well-known ultra-cool dwarfs
observed with ground telescopes. These results provide a pre-launch
approximation to the expected accuracy of the Gaia parameter estimates
(mainly for the effective temperature). Finally, Sect.
\ref{conclusions} summarises the main results of this work and
describes the experiments that are currently being carried out to complete this study.

\section{Gaia sample of ultra-cool dwarfs \label{gaia}}

\subsection{Brief description of the {\Gaia} capabilities}
The {\Gaia} astrometric mission was approved by the European Space
Agency in 2000 and the construction of the spacecraft and payload is
on-going for a launch in mid 2013. {\Gaia} will continuously scan the
entire sky for five years, yielding positional and velocity measurements
with the accuracies needed to produce a stereoscopic and kinematic
census of about one billion stars throughout our Galaxy and beyond.
The stellar survey will be complete to {\Gaia} magnitude $G =
20$ mag, with a precision of 24 $\mu$as at magnitude V=15 for a
solar-type star (G2V). {\Gaia} will be equipped with two
spectrophotometers operating in the 330-680 {\rm nm} range (blue
photometer or BP) and in the 640-1000 {\rm nm} range (red photometer
or RP). Both spectrophotometers are based on a dispersive-prism
approach, and the spectral resolutions are, in both cases, wavelength
dependent. The RP has a varying resolution from 7 nm pixel$^{-1}$ at
640 nm to 15 nm pixel$^{-1}$ at 1000 nm, while the BP photometer
resolution reaches from 4 to 32 nm pixel$^{-1}$ in its wavelength
range. The details of the Gaia passband G and the photometric
performances of the {\Gaia} instruments are summarised in
\cite{2010A&A...523A..48J}.

\subsection{Ultra-cool dwarfs with Gaia}

\begin{figure*}[thb]
  \centering
  \includegraphics[scale=0.6]{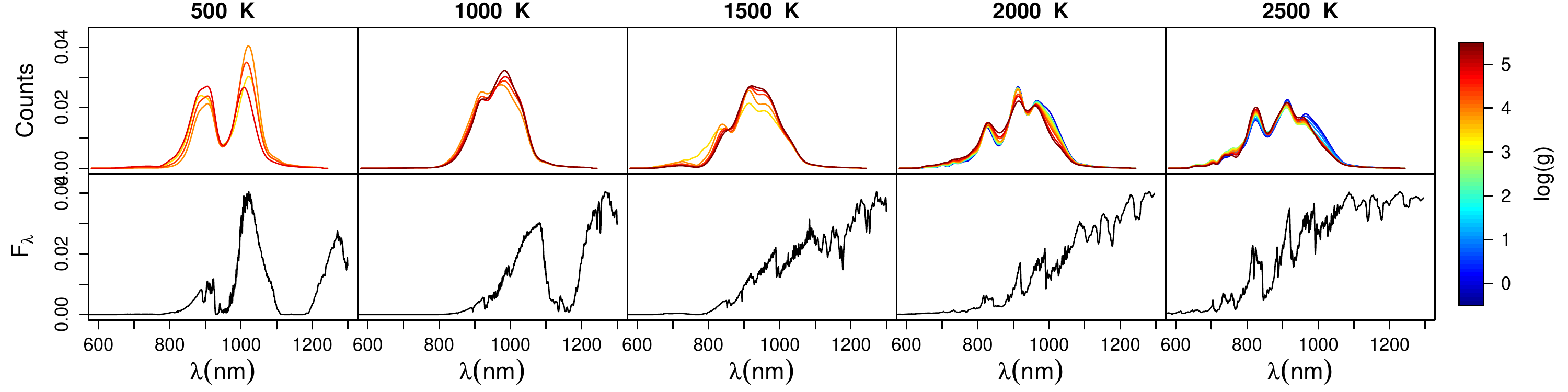}
  \caption{Normalised sample spectra from the BT-Settl library. The
    top row shows simulated {\Gaia} RP spectra of BT-Settl models for
    {\teff} = 500, 1000, 1500, 2000, and 2500 K. The vertical
      axis is proportional to the number of photons detected in each
      wavelength bin. The line colours reflect the various values of
    $\log(g)$ available in the library of models according to the
    colour scale on the right. The bottom row shows the original
    spectra with the same temperatures as in the top row,
    $\log(g)=5.0$ and solar metallicity.}
  \label{models-gog}
\end{figure*}

The spectral energy distributions (SEDs) of ultra-cool dwarfs all peak
in the infrared range, and we do not expect any significant flux in
the BP range (see Fig. \ref{models-gog}). Therefore, we will mainly be
concerned with detecting and characterising of these stars using
RP spectra.

A rough estimate of the number of ultra-cool dwarfs that will be
detected by {\Gaia} per bin of spectral type can be obtained asuming a
local volume density such as the one compiled by
\cite{2008A&A...488..181C}. These authors compiled (or
derived) local densities of late-type stars and
brown dwarfs between spectral types M3 and T8 from the literature, to provide
estimates of contamination rates by these objecs in deep photometric
surveys that searched for substellar objects or high-redshift quasars. The
overall shape of the local density as a function of spectral type
shows three local maxima at M3, L5, and T8, and reaches a
minimum of 0.22$\times 10^{-3}$ objects per cubic parsec at spectral
types T0-T1.  For each spectral type, we computed the distance at
which a main-sequence object of the corresponding I-band absolute
magnitude would reach the {\Gaia} detection limit of G=20 using the
Gaia object generator. This (hereafter GOG) is
one of the three {\Gaia} generators of simulated data that also
include GASS (telemetry generator) and GIBIS (image generator). The GOG is
a tool designed to obtain directly simulated catalogue and main
database data for the {\Gaia} satellite, passing through the entire
mission data reduction chain \citep{2012arXiv1202.0132R}. The outputs
are astrometric, photometric, and spectroscopic epoch and final
data. To simulate the main database data lifecycle, GOG uses error
models whose formulas are coded using the current knowledge of the
{\Gaia} mission performances. In this work, we used GOG
simulations of two synthetic libraries of ultra-cool dwarf spectra to derive the maximum distance at which an ultra-cool dwarf can
be detected as a function of the I-band absolute magnitude, and other
properties of the {\Gaia} UCD sample.

\begin{figure}[thb]
  \centering
  \includegraphics[scale=0.4]{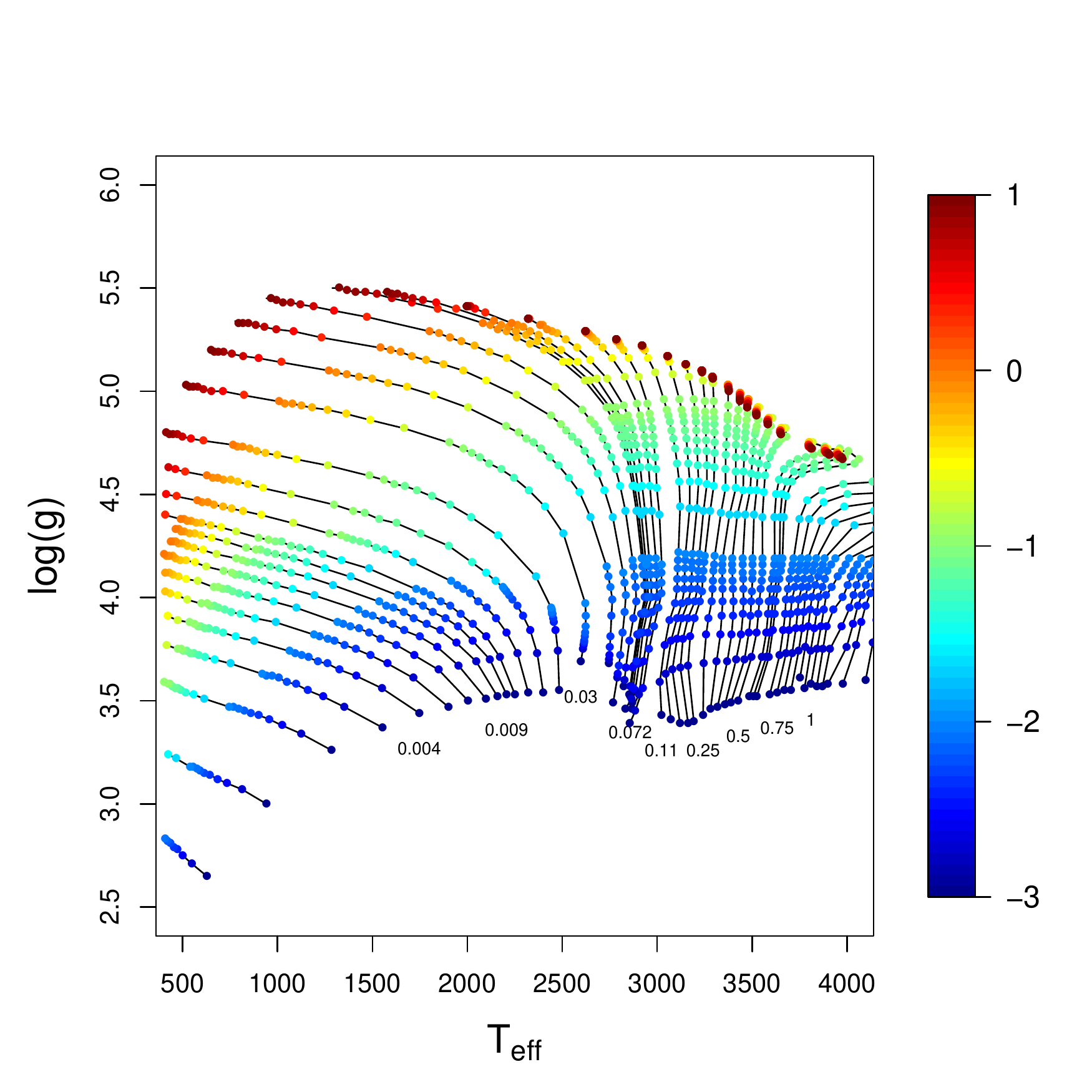}
  \caption{Evolutionary tracks in the {\teff}-{\logg} space for
    ultra-cool dwarfs according to the BT-Settl library. Each line
    corresponds to a different mass in the range
    0.0005--1.4M$_{\odot}$ as labelled below selected tracks. Filled
    circles represent individual models in the grid. These are
    coloured according to the decimal logarithm of the age in
    Gigayears as indicated by the colour scale on the right. Effective
    temperatures are measured in Kelvin and gravities in cm
    s$^{-2}$.}
  \label{evolutionary-tracks}
\end{figure}

The first library is a composite of the AMES-Cond and AMES-Dusty
models described in \cite{2001ApJ...556..357A}. The validity ranges
for these models are {\teff} $<$ 1400 K (AMES-Cond) and {\teff} $>$
1700 K (AMES-Dusty). Therefore, there is a gap in the validity (not in
the coverage) in effective temperature between 1400 and 1700
models. Models in the interregnum are available in both model families
(and hence, no gap in coverage exists), and will be used in this work
to interpolate between the validity domains. The second library is the
BT-Settl family of models \citep{Allard12}, valid across
the entire range of effective temperatures. Figure
\ref{evolutionary-tracks} shows the evolutionary tracks for the
BT-Settl library in the {\teff}-{\logg} ~space for a range of masses
between 0.0005 and 1.4 M$_{\odot}$. Throughout this work we
measure effective temperatures in Kelvin and gravities in cm
s$^{-2}$. The simulation of the synthetic spectra is carried out
  in practice by the so-called coordination unit 2 (CU2) of the
{\Gaia} DPAC. In most of
this work we concentrated the results obtained with the
BT-Settl library of models which produced better fits to the observed
spectra used for validation in Sect. \ref{valid}.

These two model libraries will also be used to define the mapping
between the source parameters and {\Gaia} observations described in
section \ref{method}. Figure \ref{models-gog} shows a sample of
spectra from the BT-Settl library together with the simulated {\Gaia}
RP spectra for a range of temperatures between 500 K and 2500 K. We
obtained Gaia simulated data using the GOG. In the
simulation of the libraries, CU2 takes into account the evolutionary
tracks provided with the models.

\begin{figure}[thb]
  \centering
  \includegraphics[scale=0.4]{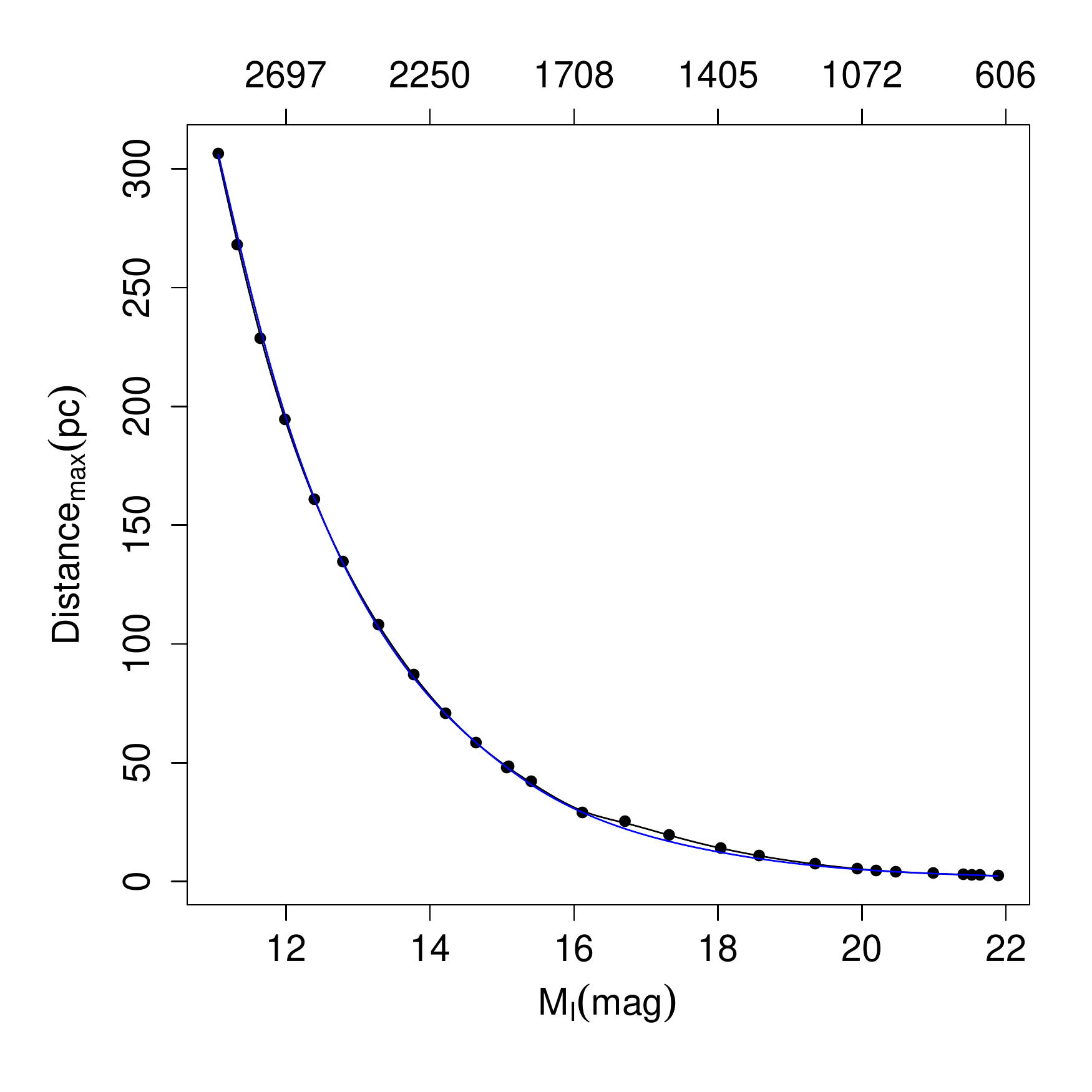}
  \caption{Maximum distances at which an ultra-cool dwarf can be
    detected by {\Gaia} at the limiting magnitud G=20 as a function of
    its absolute magnitude in the I band. These have been derived from
    BT-Settl models (filled circles) and the continuous lines
    represent the interpolation used in deriving the expected
    counts per spectral type bin in Table \ref{predcounts-table}. The
    black continuous line corresponds to {\logg}=5.0 and the blue
    line to {\logg}=3.5. The top axis shows the effective temperature
    measured in Kelvin for a {\logg}=5.0 object with the
    absolute I magnitudes shown in the $x$ axis, according to the
    BT-Settl models. The {\teff}--M$_{\rm I}$ mapping is only
    bi-valued below 600 K.}
  \label{maxdist}
\end{figure}

\begin{figure*}[thb]
  \centering
  \subfigure[]{\label{predcounts-plot}
  \includegraphics[scale=0.35]{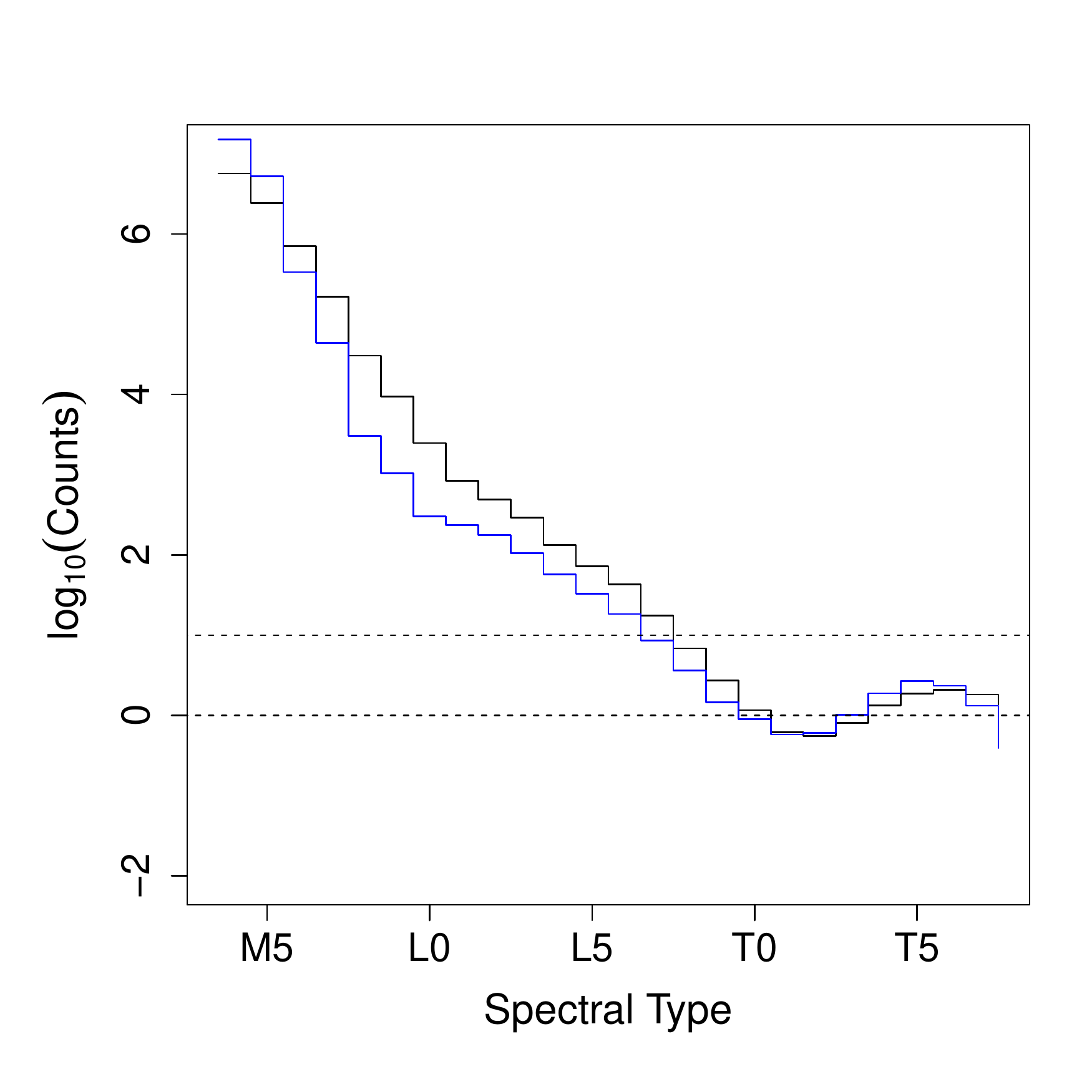}}
  \subfigure[]{\label{predcountsG-plot}
  \includegraphics[scale=0.35]{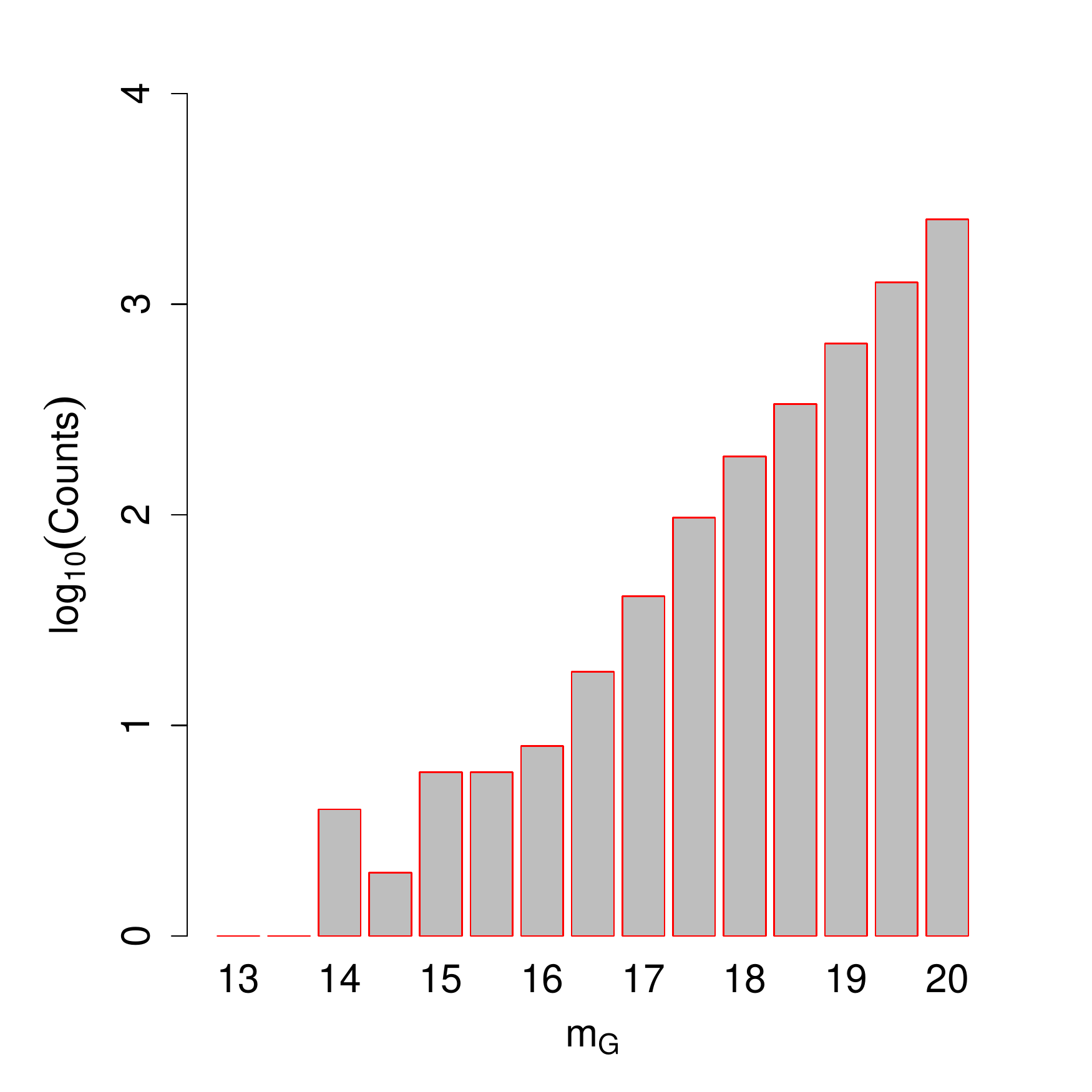}}

  \caption{Predicted number of counts per spectral type bin (a) and
    per apparent G magnitude (b), in logarithmic units.  The black
    line in the left panel corresponds to the derivation based on the
    relation between spectral type and I-band absolute magnitude
    included in \cite{2008A&A...488..181C}, while the blue continuous
    line corresponds to the relation derived from the BT-Settl model
    family and the SLC calibration. The two horizontal (dashed) lines
    indicate the levels of predicted counts equal to one and ten. The
    right-hand side plot has been obtained assuming the relation
    between effective temperature and I-band absolute magnitude
    derived from the BT-Settl models and the SLC calibration.}
\end{figure*}

Figure \ref{maxdist} shows the maximum distances to main-sequence UCDs
as a function of their I-band absolute magnitude. The expected number
of detectable main-sequence objects in each spectral type bin can be
estimated multiplying a volume density estimate \cite[in our case, the
  one in][ for spectral types between M3 and T8]{2008A&A...488..181C}
by the volume of a sphere with a radius equal to the maximum distance
at which an UCD corresponding to the spectral type under consideration
can be detected (assuming solar metallicity). In computing
these expected number counts per spectral type bin, we need to
define a relationship between absolute I-band magnitude and spectral
type. We used two such relations. The first one is included in
Table 3 of \cite{2008A&A...488..181C} (and reproduced in Table
\ref{predcounts-table} for convenience). The second relationship is
derived from the I-band magnitudes and effective temperatures of the
BT-Settl model family, combined with the calibration of effective
temperatures with spectral types by \cite{2009ApJ...702..154S}. We used
the analytic formula based on optical spectral types of L dwarfs and
infrared spectral types of the T dwarfs. This calibration (hereafter
refered to as SLC calibration) is valid in the M6 to T8 range. This
analysis results in the values listed in Table \ref{predcounts-table}
and illustrated in Fig. \ref{predcounts-plot}. Table
\ref{predcounts-table} lists the relationship between spectral type
and absolute magnitude in the I-band given in
\cite{2008A&A...488..181C} in column 2 for reference. This results in
the estimated number of counts under the column header
$Counts_{CBK08}$. The relationship between spectral type and absolute
magnitude in the I-band implicit to the BT-Settl model family is
included in column 3. This results in the estimated number of counts
under the column header $Counts_{BT-Settl}$. The SLC spectral
type--effective temperature calibration (see Sect. \ref{valid}) is
included in column 6, and the (G$_{\rm BP}$-G$_{\rm RP}$) colour index
computed from the noiseless GOG simulations of BT-Settl model
atmospheres is included in column 7. Since the volume densities tabulated in
\cite{2008A&A...488..181C} refer only to the main sequence, these
estimates do not take into account the potential detection of low-gravity objects (i.e., essentially very young objects). The final
increase in the expected number of counts is due to the steep increase
in the volume density for spectral types later than T0. In
deriving these estimates we only used BT-Settl models of solar
metallicity. The expected number of detections per apparent G
magnitude is shown in Fig. \ref{predcountsG-plot}.

\begin{table*}[thb]
\caption{ Predicted number of counts assuming solar metallicy and
    auxiliary relations.}
\label{predcounts-table} 
\centering      
\begin{tabular}{c c c c c c c} 
\hline\hline             
spectral type & $M_{\rm I,CBK08 }$ & $M_{\rm I, BT-Settl}$ & $Counts_{\rm CBK08}$    &$Counts_{\rm BT-Settl}$    & {\teff} & (G$_{\rm BP}$-G$_{\rm RP}$)\\ 
\hline                       
M3--4\,V	&  9.07  & 10.24  &  $1.6\times 10^{7}$   &  $5.7\times 10^{6}$   &  3269  & 2.86   \\ %
M4--5\,V 	& 10.14  & 10.86  &  $5.4\times 10^{6}$   &  $2.4\times 10^{6}$   &  3065  & 3.33   \\ %
M5--6\,V  	& 11.91  & 11.39  &  $3.4\times 10^{5}$   &  $7.0\times 10^{5}$   &  2889  & 3.74   \\ %
M6--7\,V  	& 12.90  & 11.92  &  $4.3\times 10^{4}$   &  $1.7\times 10^{5}$   &  2731  & 4.10   \\ %
M7--8\,V  	& 14.16  & 12.47  &  $3.0\times 10^{3}$   &  $3.0\times 10^{4}$   &  2583  & 4.44   \\ %
M8--9\,V  	& 14.68  & 13.05  &  $1.0\times 10^{3}$   &  $9.4\times 10^{3}$   &  2441  & 4.77   \\ %
M9--L0\,V 	& 15.23  & 13.66  &  $3.0\times 10^{2}$   &  $2.5\times 10^{3}$   &  2303  & 5.03   \\ %
L0--1\,V  	& 15.27  & 14.30  &  $2.3\times 10^{2}$   &  $8.4\times 10^{2}$   &  2166  & 5.07   \\ %
L1--2\,V  	& 15.69  & 14.95  &  $1.7\times 10^{2}$   &  $4.9\times 10^{2}$   &  2032  & 5.10   \\ %
L2--3\,V  	& 16.19  & 15.44  &  $1.0\times 10^{2}$   &  $2.9\times 10^{2}$   &  1902  & 5.62   \\ %
L3--4\,V  	& 16.67  & 15.87  &  $5.4\times 10^{1}$   &  $1.3\times 10^{2}$   &  1777  & 5.35   \\ %
L4--5\,V  	& 17.14  & 16.38  &  $3.2\times 10^{1}$   &  $7.2\times 10^{1}$   &  1660  & 5.15   \\ %
L5--6\,V  	& 17.61  & 16.97  &  $1.8\times 10^{1}$   &  $4.3\times 10^{1}$   &  1554  & 8.02   \\ %
L6--7\,V  	& 18.07  & 17.57  &  $8.7$               &  $1.8\times 10^{1}$   &  1460  & 9.05   \\ %
L7--8\,V  	& 18.52  & 18.09  &  $3.7$               &   $6.8$               &  1381  & 9.85   \\ %
L8--9\,V  	& 18.90  & 18.49  &  $1.5$               &   $2.7$               &  1318  & 10.38  \\ %
L9--T0\,V 	& 18.95  & 18.78  &  $9.4\times 10^{-1}$  &  $1.1$                &  1270 &  10.81  \\ %
T0--1\,V  	& 19.01  & 18.97  &  $6.1\times 10^{-1}$  &  $6.1\times 10^{-1}$   &  1237 &  11.17  \\ %
T1--2\,V  	& 19.04  & 19.10  &  $6.4\times 10^{-1}$  &  $5.5\times 10^{-1}$   &  1214 &  11.44  \\ %
T2--3\,V  	& 19.04  & 19.19  &  $1.1$               &   $8.0\times 10^{-1}$  &  1196 &  11.64 \\ %
T3--4\,V  	& 19.07  & 19.29  &  $2.0$               &   $1.3$               &  1176 &  11.84 \\ %
T4--5\,V  	& 19.22  & 19.44  &  $2.9$               &   $1.9$               &  1144 &  12.15 \\ %
T5--6\,V  	& 19.61  & 19.68  &  $2.6$               &   $2.1$               &  1086 &  12.61 \\ %
T6--7\,V  	& 20.35  & 20.05  &  $1.2$               &   $1.8$               &  986  &  12.99 \\ %
T7--8\,V  	& 21.60  & 20.62  &  $4.0\times 10^{-1}$  &   $1.4$               &  824  &  12.46   \\ %
\hline                                
\end{tabular}                         
\end{table*}

\subsection{Selection criteria and contamination rates}

In the {\Gaia} processing pipeline a source will only be characterised as
an ultra-cool dwarf if the following conditions are
met:
\begin{enumerate}
\item Its estimated distance is smaller than the maximum distance at
  which an ultra-cool star can be detected. This is the distance at
  which the brightest ultra-cool star would have a {\Gaia} limiting
  magnitude of G=20. We estimated the maximum brightness of an
  ultra-cool star from the BT-Settl models, and it corresponds to the
  hottest model with the lowest surface gravity.
\item The source is fainter (in the G band) than the brightests model
  placed at the same distance as estimated by Gaia.
\item The BP-RP colour index is higher (redder) than the minimum
  (bluest) colour index found in the model libraries.
\item The celestial coordinates and proper motions are not consistent
  with solar system Keplerian motions.
\end{enumerate}

Sources that fulfil these criteria (within some margin that depends
on the measurement uncertainties) are subsequently processed
to estimate their effective temperature and surface gravity as
described in section \ref{method}. Objects not detected in the BP band
(but fulfilling all other criteria) will nevertheless be selected because the non-detection is itself an indication of a red
spectrum. This could imply a potential contamination of the UCD sample
by faint, nearby objects close to the {\Gaia} limiting magnitude and
colour indices that are positive (thus excluding white dwarfs), but
bluer than the bluest UCD. The software that implements the
analysis of ultra-cool dwarfs distinguishes between the {\it
  selection module} that decides whether to process a given
Gaia source, and the {\it processing module} that estimates the
effective temperature and gravity of the selected sources.

To define the boundaries of brightness, colour index, and
distance that define the region of ultra-cool sources, we analysed the
GOG simulations of the AMES-Cond, AMES-Dusty and BT-Settl model
libraries (although the latter is prefered, and the former have not
been implemented in the selection software). Using GOG, we found that
the most distant ultra-cool object in the library (a source with
{\teff} = 2500 K and {\logg} = 3.5 at G=20) corresponds to a distance
of 373 pc. Thus, the first criterion limits the processing of sources
by the UCD module to sources within this radius (this is necessary optimise the processing time per source). The second criterion
examines the brightness of each detected source and compares it with
the aforementioned brightest model placed at the same distance. We
expect a strong contamination from main-sequence stars with
temperatures above the 2500 K limit because low-gravity sources are
brighter than main-sequence ones for temperatures above 1600 K. In
fact, we derived from the GOG simulation of the model libraries that a
dwarf star of $\approx 4180$ K has the same G magnitude as the lowest
gravity, hottest ultra-cool giant considered in this work ({\teff} =
2500 K, {\logg} = 3.5).  Thus, main-sequence objects up to this
temperature will be selected by the selection module according to this
criterium despite their higher temperatures (although the colour index
criterium will reject a fraction of these hotter stars; see below). We
could have used the thresholds corresponding to the main-sequence
objects, but this would have resulted in the potential loss of low-gravity
sources in the selection process. We opted for an inclusive set of
criteria in spite of the expected high contamination rates by dwarfs
hotter than 2500 K. To estimate the contamination rate (caused by the criteria but also to the measurement
errors), we conducted a numerical experiment in which we populated a one-kiloparsec cube with sources uniformly distributed in space (this is
much larger than actually needed in view of the estimated maximum
distance to an UCD). The total number of sources in each temperature
bin was generated using a probability density function derived by
interpolating and normalising the volume densitites tabulated in
\cite{2008A&A...488..181C}. Since these correspond to main-sequence
sources, we neglected the contribution from ultra-cool giants in the
following estimates.

\begin{figure}[thb]
  \centering
  \includegraphics[scale=0.5]{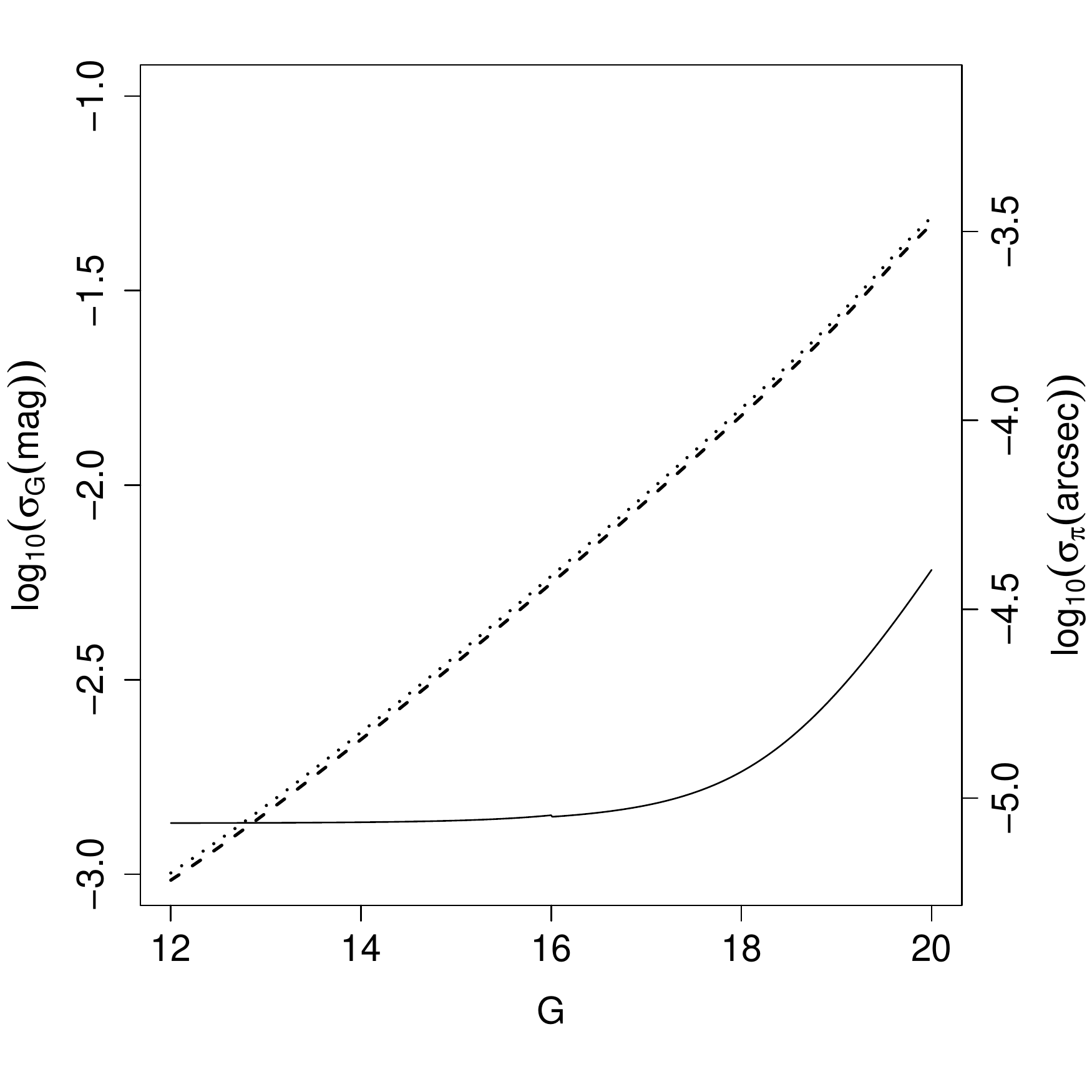}
  \caption{Current estimates of the end-of-mission uncertainties
    in the measurements of the G apparent magnitude (black line) and
    the parallax (dashed and dotted lines) as a function of the G
    apparent magnitude. The dashed line corresponds to (V-I) = 4 and
    the dotted line to (V-I) = 7.5, a plausible range for the V-I
    colour index according to the model libraries.}
  \label{sigmaGpi}
\end{figure}

For each star, and given the effective temperature and the distance to
the centre of the cube, we computed the apparent G magnitude and the
(V-I) colour index by interpolating in the BT-Settl library restricted
to dwarfs with values of {\logg} $\approx$ 5.5. We subsequently used
the current estimates of the uncertainties in the {\Gaia} magnitudes
and parallaxes (see \cite{2010A&A...523A..48J} and
\cite{debruijne2009} respectively)  to generate mock
measurements of the distances and G magnitudes by sampling from normal
distributions with the prescribed uncertainties. We did not use a
full covariance matrix since none was available at the time of
writing. We show in Fig. \ref{sigmaGpi} the error models that
we used in the simulations. The uncertainty estimate in
the G magnitude is based on Eq. 6 in \cite{2010A&A...523A..48J} and on 
the mission parameters available at the time of writing. We included a
calibration error $\sigma_{cal}=$30 mmag for a single transit.  This
value is only an educated guess since the final $\sigma_{cal}$ can
only be estimated during the operational phase. The uncertainty in the
parallax is based on Eq. 1 in \cite{debruijne2009} (see \cite{db2012}
for a more recent review without analytic expressions for the
astrometric uncertainty).

The resulting sample of stars in the kiloparsec cube was then examined
to determine the properties of the subsample that fulfilled
the criteria enumerated above (except for the non-Keplerian
motion). We found that none of the ultra-cool dwarfs thus generated is
missed in the selection process due to errors in the
measurements. This was expected given i) the inclusiveness of the
criteria, ii) the relative proximity of these faint sources to the
Sun, and iii) the fact that we only generated main-sequence objects in
the simulation (and not low-gravity young sources that would be
closer to the selection thresholds; this question will be re-examined
during the software validation phase with real {\Gaia} data and may
result in more inclusive thresholds if we identify examples of this
kind missed by the selection module). The uncertainties in the G
magnitude and colour index typical of UCDs according to the
simulations are negligible in this context.

\begin{figure}[thb]
  \centering
  \includegraphics[scale=0.35]{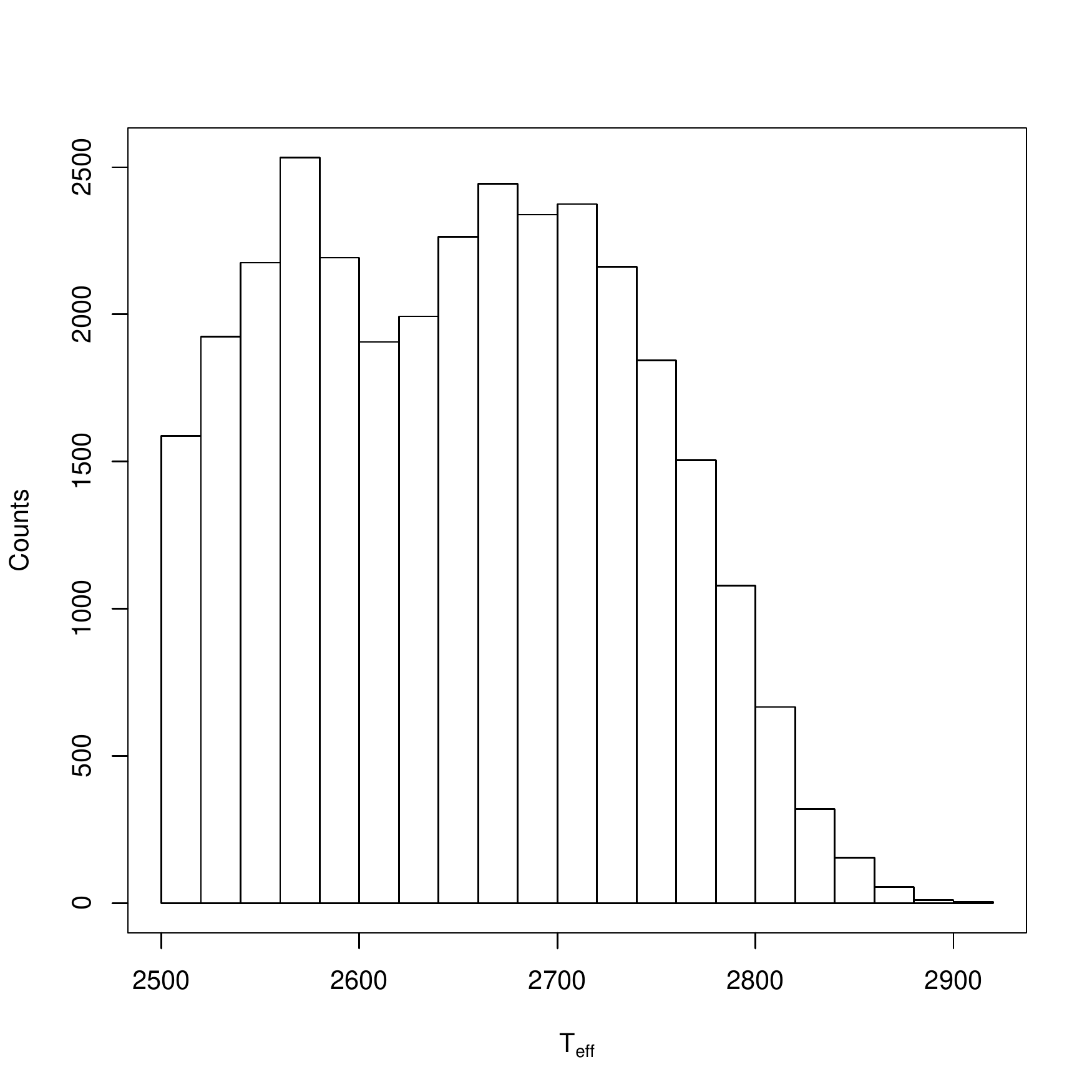}
  \caption{Histogram of the temperatures of non-UCD stars that pass
    the selection criteria of the {\Gaia} UCD module.}
  \label{contam-hot}
\end{figure}

The (G$_{\rm BP}$-G$_{\rm RP}$) colour index thresholds in the
selection criteria are defined by the bluest model in the GOG
simulations of the model libraries. For the BT-Settl library of models
and metallicities [M/H] between -2 and 0.5, we find that the
(G$_{\rm BP}$-G$_{\rm RP}$) colour index is in the 4.1--14.5
range. The distribution of effective temperatures in the resulting
sample of non-UCD stars that fulfil the criteria is shown in
Fig. \ref{contam-hot}. The shape is determined by the combination of
the various selection criteria and the uncertainties in the
measurements.

Fig. \ref{contam-hot} implies a contamination rate of abound 85\%. This
is explained by the exponential increase in volume densities for the
mid- to late-M spectral types. To take this high
contamination rate into account, we
generated regression models in Sect. \ref{method} that are capable of predicting effective
temperatures up to 4000 K, and not only in the UCD regime below 2500
K. Therefore, even if the selection module selects these contaminants,
the processing module will allow for the obtention of purer samples of
UCDs by filtering out objects with estimated temperatures above a
given threshold.

In computing the contamination rate we did not consider
extra-galactic contaminants because their parallax measurements would
be incompatible with the selection criteria defined above.


\section{Methodology \label{method}}

As mentioned in Sect. \ref{intro}, {\Gaia} is expected to detect and
process a number of sources close to one billion. This order of
magnitude necessitates automating the detection and
characterization processes. The main objective of the eighth
coordination unit of the {\Gaia} DPAC is to implement and evaluate
automated procedures to derive the astrophysical parameters of the
sources detected by the {\Gaia} instruments. This necessarily involves
techniques developed in the fields of statistical learning and data
mining. In the framework of statistical learning methods, this is
accomplished with regression models that are constructed from sets of
examples (the so-called training set) that link the independent
variables (the {\Gaia} observations in our case) with the dependent
variables that we wish to infer. In this section we describe several
techniques taken from the field of statistical learning, aimed at
providing a reproducible, systematic characterization of the {\Gaia}
UCD candidate sources. 

In collecting the aforementioned sets of examples, libraries of
stellar models and associated synthetic spectra provide a homogeneous
and consistent set of examples that uniformly cover  the parameter
spaces under consideration. In contrast, catalogues of observed
spectra with associated astrophysical parameters derived from them
tend to be fragmentary in nature: each collection covers only a
reduced range of parameters, and it is necessary to combine several
catalogues to obtain a coverage that, even then, may contain gaps and/or
insufficiently sampled regions (this is especially true for the
parameter {\logg}, which is missing from most compilations). Each
catalogue has its own observational setup, systematic errors, and
selection effects. Also, observational biases may favour the abundance
of examples in particular regions of the parameter space, which then
translate into systematic biases in the predictions. Furthermore,
parameter estimation across catalogues can be inconsistent, and we
indeed find slightly different spectral types
assigned to the same source in different catalogues, which
reveals a certain degree of subjectivity in the assignements. With all
these considerations in mind (reproducibility, consistency,
homogeneity, and uniformity of the training set), we prefered to
construct our regression models from the aforementioned synthetic
libraries. These libraries parameterise the models with physical
magnitudes (effective temperatures, gravities, and metallicities) and
not with spectral types.

Model libraries are nevertheless imperfect in the sense that they do
not reproduce each and every spectral feature in the real spectra of
UCDs or its exact dependence on the physical parameters. The problem
of the potential mismatch between stellar model libraries and
observations appears whenever physical parameters are to be produced,
beyond phenomenological descriptions such as spectral types. Spectral
types can be inferred without the need for synthetic models, while
going from spectra (or spectral types) to physical parameters requires
model libraries to interpret them. In this work we do not attempt to
build regression models to infer spectral types. Given the low
spectral resolution of the {\Gaia} red spectrophotometer (RP), and its
spectral coverage, most of the spectral features used to decide the
spectral type remain unresolved or unobserved, and thus, spectral
types derived from them would be of limited use.

The systematic effects of the selection of model libraries can be
characterised to some extent by comparing the predictions from
different model families like those presented in the previous
section (AMES-Cond, AMES-Dusty, and BT-settl). In Sect. \ref{valid} we
also attempt to validate our models with an external set of effective
temperatures derived from ground-based observations and spectral types
via a calibration that, inevitably, encompasses another synthetic
stellar library.  The two families of models introduced in the
previous section (AMES-Cond and -Dusty, and the BT-Settl models) are
used to define the relationship between {\Gaia} observables (the RP
spectrum) and the parameters that we intend to estimate, namely
{\teff} and {\logg}. This relationship is captured in a regression
model (not to be mistaken for the physical models of the stellar
atmospheres and the resulting synthetic spectra) constructed from the
set of examples defined by the two synthetic spectral libraries. Each
spectrum in the libraries (e.g. those represented in the lower row of
Fig. \ref{models-gog}) is characterised by the set of physical
parameters that identifies the stellar atmosphere used to synthesize
it. This set of examples composed of the spectrum plus the
corresponding physical parameters is referred to in the following as
the training set. The set of independent examples used to assess the
accuracy of the models will be referred to as the test set.

As mentioned before, the training set is constructed using the
AMES-Cond, -Dusty, and BT-Settl model libraries (restricted to
  effective temperatures below 4000 K)) and simulating {\Gaia}
observations of these synthetic models using GOG (version
  8.0). {\Gaia} has a nominal duration of five years. During this
period, a source is observed on average approximately 70 times. Each
time a source is observed (i.e., each transit), an epoch RP spectrum is
obtained. The characteristics of this spectrum depend on the
instrument design including prism and CCDs, and the transit geometric
properties. In general, we can assume that the spectrum will be spread
along 60 spectral bands with a non-uniform dispersion. In each
transit, the position of the source (continuous) spectrum may vary
with respect to the discrete CCD pixel array depending on the transit
geometric details. This transit dependence results in different
spectra obtained for each epoch, because the wavelength coverage of
each pixel is different. This subresolution information can then be
used to produce oversampled combined spectra. In Fig. \ref{models-gog}
and in the experiments carried out in this work that are described in the
following sections we assumed an end-of-mission oversampling factor of
three, resulting in a total number of flux bins of 180.

\subsection{Regression models}

In this section we describe three different types of
statistical regression models: $k$-nearest
neighbours \citep{Cover:Hart:1967}, Gaussian processes \citep[see
  e.g.][]{Bishop:2006:PRM:1162264,Rasmussen2006gaussian}, and Bayesian
inference \citep[see e.g.][]{sivia2006data}. A previous analysis also
included support vector machines
\citep{Vapnik:1995:NSL:211359,Cortes:Vapnik:95} and kernel partial
least squares \citep[KPLS, ][]{Rosipal01kernelpartial}.

$k-$nearest neighbours estimation (kNN) is by far the simplest model
and derives the parameter values as the weighted average of the
elements in the training set that are closest to the input spectrum in
a given metric. In our case, the euclidean distance is used to define
proximity and the weights are defined as the inverse of this
distance. A maximum-likelihood estimate of the uncertainties in the
estimated parameters can be computed by modelling the distribution of
inverse distances to the nearest neighbours (under the assumptions
that the model grid is sufficiently dense and the neighbourhood has
the appropriate size to sample a unimodal likelihood). It has the
disadvantage that the full training set has to be stored and accessed
each time that the regression model is used to predict the physical
parameters ({\teff} and {\logg}) of a source, and furthermore, it is
severely affected by the so-called curse of dimensionality. This is reflected in the need for exponentially growing
training set sizes as the problem input dimensionality increases. The
exponential growth is required to ensure that the nearest neighbours
are sufficiently close to provide an accurate estimate of the
parameters.

Support vector machines (SVMs) and Gaussian processess (GPs) are two
examples of kernel methods. These methods transform the regression
problem into a dual representation where the constituents of the model are 
no longer the input features (or nonlinear mappings thereof) but
their scalar products expressed as kernel functions. More details of
this dual representation can be found in the textbooks by
\cite{Bishop:2006:PRM:1162264} and \cite{Hast:Tibs:Frie:2001}.

A GP is defined as a probability distribution over
functions such that the joint probability of the random variables
defined by their evaluations at a certain set of input vectors (the
training set in our case) is Gaussian. If we assume that the nature of
our problem is such that the probabilistic distribution for both the
targets of our training set and the {\Gaia} observations to be
characterised is well captured by a multivariate Gaussian
distribution, we can construct the model by computing its
covariance matrix. It turns out that we can calculate it by evaluating
the kernel functions at the input vectors of the training set. These
kernel functions encode both the assumed error model for our
determination of the target values in the training set and the length
scale for the correlations between the examples in the training
set. In the full Bayesian treatment these two parameters can be
considered hyperparameters and are marginalised out. Here we have
determined optimal values for them through exhaustive cross-validation
experiments (see below).  For the SVM,
the model representation is sparse in the sense that it only depends
on kernel evaluations on a small subset of the training set (the
support vectors). In regression problems, the support vectors are
those that lie within the boundaries of, or outside a so-called
insensitive tube. Support vector regression involves the search for best (minimum error) solution in a space of hyperparameters,
similar in nature to those discussed for GPs.

In this section we describe a parallel aproach to parameter estimation
based on Bayes' theorem. The main advantage of Bayesian parameter
estimation stems from the fact that it provides not only an estimate
of the parameters but also a full multivariate probability density
distribution for the set of parameters given the observations.

We assume that we observe the spectrum \( \mathbf{s} \) of an
ultra-cool dwarf. In the Bayesian framework, we seek the probability
density function (PDF) of the physical parameters of the UCD, given
this spectrum \( \mathbf{s} \). Bayes' theorem provides this PDF as

\begin{equation}\label{eq:Bayes}
	\pi (\bm{ \theta } | \mathbf{s}) = \frac{\pi (\bm{ \theta })\cdot
          f_{\bm{ \theta }} (\mathbf{s})}{m(\mathbf{s})},
\end{equation}

where \(\bm{ \theta }= (T_{\rm eff}, \log (g)) \) is the vector of
parameters that we intend to derive, $\pi (\bm{ \theta })$ is the
prior probability distribution of these parameters, and $f_{\bm{ \theta
}} (\mathbf{s})$ is the likelihood of the spectrum given the
parameters. 

The denominator in the right-hand side of Eq. \ref{eq:Bayes} \(
m(\mathbf{s}) \) is the \emph{prior predictive distribution} or
\emph{evidence} defined as
\begin{equation}\label{eq:evidence1}
 	m(\mathbf{s}) = \int f_{\bm{ \theta }}
        (\mathbf{s})\,\pi (\bm{ \theta})\,d\bm{ \theta},
\end{equation}
and can be viewed as a normalisation constant.

The likelihood term encompasses a predictive model for the spectra
given the parameters, together with a probabilistic error model. In
our case, we used the two stellar libraries mentioned above
(COND+DUSTY and BT-Settl) to build the predictive model. This is
captured in two three-layer perceptrons (a kind of neural network)
each trained with the stellar spectra of the corresponding
library. The neural network captures a multivariate regression model
that can be used to perform an interpolation. Extensive experiments to
derive the optimal network architecture result in hidden layers of 20
hidden units for the two model libraries. With it, we can generate
output synthetic spectra for any input value of \(\bm{ \theta }\)
within the grid boundaries. The neural network exactly predicts the
spectra in the COND+DUSTY and BT-Settl grids, and smoothly interpolates for intermediate values. The unique flux predicted by the
neural network for each wavelength in the spectrum can be turned into
a probabilistic statement by convolving it with the predicted
measurement errors in the current {\Gaia} model. At present, the error
model for the RP spectra consists of a Gaussian distribution with a
covariance $\Sigma$ that depends on a series of instrument parameters,
on the number of transits, and on the flux itself (see
\cite{2010A&A...523A..48J}, discussed above). Thus,
\begin{equation}\label{eq:evidence2}
 	f_{\bm{ \theta }}(\mathbf{s})=f_{(\mathbf{x_{\theta}}, \Sigma)}(\mathbf{s}),
\end{equation}
with \(\mathbf{x_{\theta}}\) being the prediction from the neural
network for the parameter set $\theta$, and \(
f_{(\mathbf{x_{\theta}}, \Sigma)}(\mathbf{s}) \), the probability
density function of a normal distribution \( \mathcal{N}
(\mathbf{x_{\theta}}, \Sigma)\) evaluated at the observed spectrum
$\mathbf{s}$.

{\em Nested sampling}~\citep{Skilling:2006fk} is a Monte Carlo procedure
used to calculate \( m(\mathbf{s}) \) that can extract samples from \(
\pi (\bm{ \theta } | \mathbf{s}) \) as a by-product of these
calculations. The algorithm exploits the relation between the
likelihood \( f_{\bm{ \theta }}(\mathbf{s}) \) and the \emph{prior
  volume} \( X \) defined by
\begin{equation}\label{eq:PriorVolume}
	X(\lambda) = \int_{f_{\bm{ \theta }} (\mathbf{s}) > \lambda}
        \pi (\bm{ \theta }) \,d\bm{ \theta },
\end{equation}
i.e., the volume of \( \pi (\bm{ \theta }) \) over the region of
parameter space contained within the $\lambda$ likelihood iso-contour
\( f_{\bm{ \theta }}(\mathbf{s}) = \lambda \). Our prior density \(
\pi (\cdot) \) can be set in two ways: i) according to a bidimensional
uniform distribution in \( (400, 3500) \times (3.5, 5.5) \), or ii)
using a physical prior based on the temperature histograms discussed
in the previous section, and a {\logg} prior that favours values
typical for the field main-sequence dwarfs. We here studied the
predictions of a Bayesian module for regression based on a flat prior
in both parameters {\teff} and {\logg}. This represents the
framework for the analysis of the influence of physical priors (based
for example on estimated detection rates such as those presented in
section \ref{gaia} or on the astrometric and photometric data provided
by {\Gaia}) to be included in a subsequent paper in this series.

We used \emph{ellipsoidal sampling}~\citep{Shaw:2007uq} to estimate the
parameters \( T_{\rm eff} \) and \( \log (g) \). This method is a
variant of {\em nested sampling} that approximates the iso-likelihood
of the point to be replaced in the nested sampling step by an \( n
\)-dimensional ellipsoid (\( \epsilon_N \)) derived from the
covariance matrix of the current active points (see
\cite{Skilling:2006fk} for more details). Since our stellar models are
restricted in parameter space to the region \( 500 \leq T_{\rm eff}
\leq 3500\), \(3.5 \leq \log g \leq 5.5 \), points drawn from \(
\epsilon_N \) but outside of the parameter space will be discarded.

Even though ellipsoidal sampling is aimed at simplifying the
computation of the evidence, it is also possible to derive posterior
probabilities from it as a by-product. Once the algorithm has
converged, the resulting sample can be interpreted as a sample from
the posterior probability if we weight the importance of each point by
a factor $p_i$ defined as

\begin{equation}\label{eq:weights}
p_i = \frac{f_{\bm{ \theta}_i} (\mathbf{s}) \cdot w_i }{m(\mathbf{s})},
\end{equation}

with \( w_i = 1/2 (X_{i-1} - X_{i+1} ) \). The justification for this weighting scheme is described 
in~\cite{Skilling:2006fk}. Thereafter, we can obtain summary
statistics such as the mean of the posterior probability density of
effective temperatures using the classical first-order moment of the
distribution:

\[
	\hat{T}_{\rm eff} = \sum_{i=1}^n T_{{\rm eff},i} \cdot p_i,
\]
with $T_{{\rm eff},i}$ the effective temperature value in the $i$-th
sample, and $p_i$, the weights calculated according to Eq.
\ref{eq:weights}. We use this summary statistic (the
first-order moment) to estimate $T_{\rm eff}$, and will discuss the
effect of using other alternatives such as the posterior mode in a
subsequent paper. It is planned that the {\Gaia} catalogue will contain
the samples obtained for each UCD candidate.

Ellipsoidal sampling was preferred over simpler algorithms such as {\sl
  Metropolis-Hastings} or nested sampling because tests carried out
with these algorithms and several proposal densities produced too low
acceptance rates because of the particular shape of likelihood
landscape. This results in posterior samples that are not
independent. These problems are more severe in on-going applications
of the software to highly multi-dimensional related problems such as
the estimation of star+disk parameters in pre-main-sequence systems.

\subsection{Preprocessing}

Before constructing the models, the spectra were normalised to
yield an area equal to 1 to have an appropriate scale of
values that is robust to noise and isolated outliers. This
normalisation removes the information relative to the integrated
energy flux that, combined with the distance measured by {\Gaia}, can
provide indications of the {\teff} and {\logg} values of the
source. This information was incorporated at a later stage, and
also in the consistency checks.

Other preprocessing steps were explored to determine their
impact on the performance of the algorithms. In particular, we tested
denoisification strategies based on wavelet decomposition and moving
averages. The GOG allows generating noisy simulated spectra for
any number of transits and apparent magnitude. In the experimental
setup we included GOG simulations for a set of four apparent G
magnitudes (G=15, 18, 19 and 20) and two number of transits (28 and
70). Seventy transits is an estimate of the average number of transits
after five years of observations (the nominal {\Gaia} lifetime),
whilst 28 transits corresponds to the average after two years of
observations, when a reliable evaluation of the algorithms can be
attempted. The number of transits strongly depends on the position of
the source on the sky \citep[see e.g.][]{2012A&A...538A..78L}.

Several wavelet bases were tried in the experiments including several
orders of the Daubechies, Coiflet, best-located and least asymmetric
wavelets. In addition to denoisification strategies, we explored two data
compression approaches: the well-known principal component analysis
\citep{Pear:PCA} and the local approach based on diffusion maps
\citep{Coifman_Lafon_2006}.  

\subsection{Internal validation of the regression models}

The experiments described in this section were carried out with
regression models trained with the so-called nominal dataset. This comprised GOG simulations of the original spectra in
the synthetic spectral libraries (restricted to effective temperatures
below 4000 K), and it corresponds to the nodes of a grid in the space
of {\teff} and {\logg}, with grid spacings of 100 K in {\teff} and 0.5
dex in {\logg}. The performance of the regression models was measured
by analysing the distribution of residual errors in the so-called
random dataset. This comprises spectra linearly
interpolated from the nominal grid at values of the physical
parameters ({\teff} and {\logg}) randomly spread in the evolutionary
tracks provided by the authors of the libraries of stellar models. The
regression models can be understood as continuous nonlinear mappings
between the 180-dimensional space of spectra and the one-dimensional
space of the parameter under consideration. The random dataset is not
an independent test set because it is derived from the nominal dataset
using multi-linear interpolation. In this sense, the performance
evaluation measures not only the ability of the regression models to
reproduce the training set, but unfortunately also the fidelity to the
multi-linear interpolation between the grid nodes. We refer to
the performance measures described in this section as internal errors,
in the sense that they measure the ability of the regression models to
reproduce the mapping between spectra and astrophysical parameters
that is inherent to the training set in the UCD domain (even if the
performance also measures the fidelity to the multi-linear
interpolation). Internal errors, thus, evaluate only the regression
model and its robustness against noise, but not the validity of the
training sets or their ability to reproduce observed spectral
features. In Sect. \ref{valid} we analyse the performance of the
regression models by applying them to observed spectra (downgraded to
the spectral resolution of the {\Gaia} instruments). These performance
measures, thus, comprise the internal errors and the ability of the
training sets to reproduce the spectra of real UCDs.

\begin{figure*}[thb]
\centering
\includegraphics[scale = 0.7]{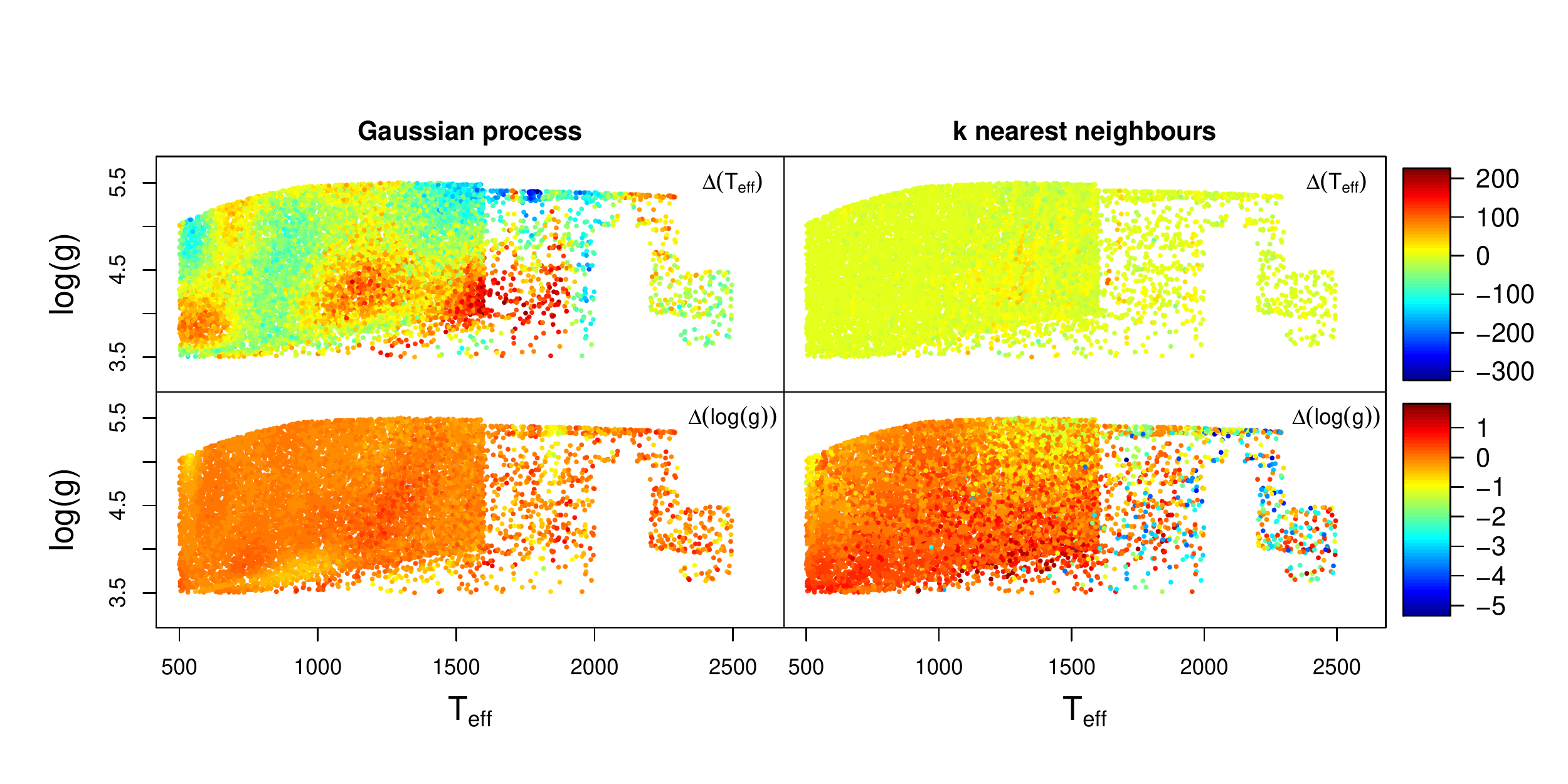}
\caption{Cross-validation errors for end-of-mission spectra of UCD
  stars at G=20. The upper panels show the errors in the {\teff}
  estimates for the GP model (left) and the $k$-nearest
  neighbours model (right). The lower panels show the corresponding
  errors in the {\logg} estimates. The colour code for the error scale
  is shown in the right-hand side of each row. Effective temperatures
  are measured in Kelvin and gravities in cm/s$^2$.}
\label{fig:cv}
\end{figure*}

The complete description and analysis of the preliminary set of
experiments described in the previous paragraphs is beyond the scope
of this article, but the conclusions derived from it can be summarised
as follows:

\begin{itemize}
\item Residual errors increase near the boundaries of the training
  sets. The region below 500 K was particularly problematic because prediction errors were unacceptably large and
  including it in the training sets also degraded the performance for
  higher temperatures. It was therefore removed from the training
  sets.
\item Denoisification is best achieved by using a moving average
  filter as measured by the root mean squared reconstruction error
  (RMSE), and the mean and median reconstruction error.
\item The lowest RMSE obtained with wavelet denoising is achieved with
  the Coiflet mother wavelet of the order of 18; similar RMSE are obtained
  with the best located wavelet of the order of 14.
\item Training the algorithms with a noisy training set adjusted to
  the signal-to-noise ratios of the test spectrum yields better
  results than training with noiseless spectra and denoisifying the
  test spectrum, regardless of the denoisification strategy (moving
  averages or wavelets).
\item Prediction errors obtained with KPLS models are significantly larger
  (RMSE systematically above 200 K for all G magnitudes and numbers of
  transits) than those obtained with GP, SVMs, or kNN.
  models;
\item Prediction errors from cross-validation experiments with the
  various models (except KPLS) and noiseless spectra are all the same
  within the experimental uncertainty as measured by the standard
  deviation of the RMSE sample obtained from ten experiments of ten-fold
  cross validation. The RMSE error for the noiseless training set is 7
  K.
\item kNN models systematically outperform both SVM and GP models when
  applied to noisy spectra in the estimation of {\teff} values. Figure
  \ref{fig:cv} shows a typical case where both the training (nominal)
  set and the test (random) set were simulated with noise
  properties for G=20 and 70 transits. The left column shows the
  distribution of residuals for the GP model ({\teff}
  and {\logg} in the upper and lower panel, respectively) and the right
  column shows the same scatter plots for the kNN model. The RMSE in
  the {\teff} predictions is 10 K for the kNN model and 62.3 for the
  GP model. The RMSE of the kNN model is unrealistically low as shown
  in the next section, and due in part to the high density of examples
  in the training set, in this context of cross-validation experiments
  to derive internal errors. In this setup, where we assumed that the
  training set of synthetic spectra reproduces the expected
  observed spectra well , the nearest neighbours are as close as allowed by
  the density of the grid of training examples (given the relatively
  high signal-to-noise ratio of the end-of-mission spectra). The
  performance of the SVM model is overall very similar to that of the
  GP model.
\item Reducing the input space dimensionality with PCA (preserving 95\% of the variance) deteriorates the model performances by up to 100\% with
  respect to the complete input spectra. This is so except for the lower signal-to-noise ratios (G=20) and in the combination of
  PCA and kNN, where
  the RMSE improves by a 30\%. The improvement decreases as the noise
  diminishes and for G=15, the two models (with and without PCA
  compression) show the same RMSE. Data compression with the
  nonlinear technique known as diffusion maps
  \citep{Coifman_Lafon_2006} results in similar performances as the
  PCA approach at a much higher computational cost.
\end{itemize}

In Sect. \ref{valid} we concentrate on the external validation
of the kNN and GP models only, because the SVM model performances are
remarkably similar to the GP models, but do not provide estimates of
the prediction uncertainty in a straightforward way.  As a result of
the previous considerations, we have several regression models based
on the $k$-nearest neighbours algorithm and GP. There
is a version of each model for the 28 and 70 transits cases, and for
the various signal-to-noise ratios that correspond to values of the G
magnitude G=15, 18, 19, and 20. For each of these cases we have a model
trained in the full input space, or on a PCA compressed version of
it. All parameters of each model, such as the number of nearest
neighbours or the kernel and noise parameters for the GP models, are
determined using ten-fold cross-validation of the models obtained in an
exhaustive exploration of the parameter space. The resulting models
are used in section \ref{valid} to analyse the expected
uncertainties when applied to spectra of ultra-cool dwarfs observed
from ground and simulated with GOG.

Unfortunately, the internal validation described above for the kNN and
GP models overestimates the real accuracy because it does not
include several systematic effects. The physical models used in
constructing of the training set (AMES-Cond, AMES-Dusty, and BT-Settl)
do not exactly reproduce all spectral features and their
correlation with the physical parameters encountered in reality.  This
may be due to several reasons, such as incomplete line lists,
inaccurate line/band opacities, or mathematical simplifications
introduced for the sake of tractability (e.g. in the dust cloud
formation and convection, in the diffusion of chemical species, in the
departures from equilibrium between the dust and gas phases, or in the
neglect of rotation). To account for these systematic
errors, a second battery of experiments is discussed in section
\ref{valid} where ground-based spectra of well-known ultra-cool stars
are degraded to the {\Gaia} resolution and are convolved with its spectral
response and error model using GOG. We aim at estimating the error
that affects each of the regression models in the various {\teff} and
{\logg} ranges covered by the empirical spectral libraries. The
spectra used for the external validation of the regression models are
described in detail in the next section.

\section{Validation with real spectra\label{valid}}

In this section, we use ground-based spectra to estimate the total
errors of the regression models. These total errors include the
internal validation errors described in the previous section, and
errors due to the imperfect representation of real spectra by
synthetic models. The ground-based spectra used for validating
the regression models are the compilations by
A. Reid\footnote{\url{http://www.stsci.edu/~inr/ultracool.html}}
\citep{2000AJ....120..473B,1997A&A...327L..25D,2000AJ....120.1085G,
  2000MNRAS.311..385G,1999ApJ...519..802K,2000AJ....120..447K,
  1999ApJ...527L.105R,2000AJ....119..369R,1999ApJ...522L..61S,
  2000ApJ...531L..61T},
S. Leggett\footnote{\url{http://staff.gemini.edu/~sleggett/LTdata.html}}
\citep{2006AJ....131.2722C,2004AJ....127.3516G,2004AJ....127.3553K},
the NIRSPEC Brown Dwarf Spectroscopic
Survey\footnote{\url{http://www.astro.ucla.edu/~mclean/BDSSarchive/#lowres}}
\citep{0004-637X-596-1-561}, and the IRTF spectral
library\footnote{\url{http://irtfweb.ifa.hawaii.edu/~spex/IRTF_Spectral_Library/index.html}}
\citep{2005ApJ...623.1115C,2009ApJS..185..289R}. None of the observed
spectra covers the full {\Gaia} wavelength range,which is especially due to
the lack of optical observations at the coolest end. Therefore, it was
necessary to complete them with models before simulating the {\Gaia}
observations of these stars for {\it ad hoc} G magnitudes (and
therefore noise levels) and total number of transits. This was
accomplished by degrading the resolution of the model spectra to that
of the observed spectrum and finding the model that yields the minimum
$\chi^2$ fit in the wavelength range of overlap restricted to the RP
passband (after removing artefacts and wavelength ranges with poor
signal-to-noise ratios). This model was then used to complete the
observed spectrum. We used all available models from the COND, DUSTY,
and BT-Settl libraries to find the best-fitting model to each of the
observations, and in all cases the BT-Settl model produced a $\chi^2$
statistic superior or comparable to the COND/DUSTY models. Spectra
with gaps, small overlap with the RP band, or very low signal-to-noise
ratios were removed from this empirical validation set to
avoid biases.

\begin{figure*}[thb]
  \centering \includegraphics[scale=0.7]{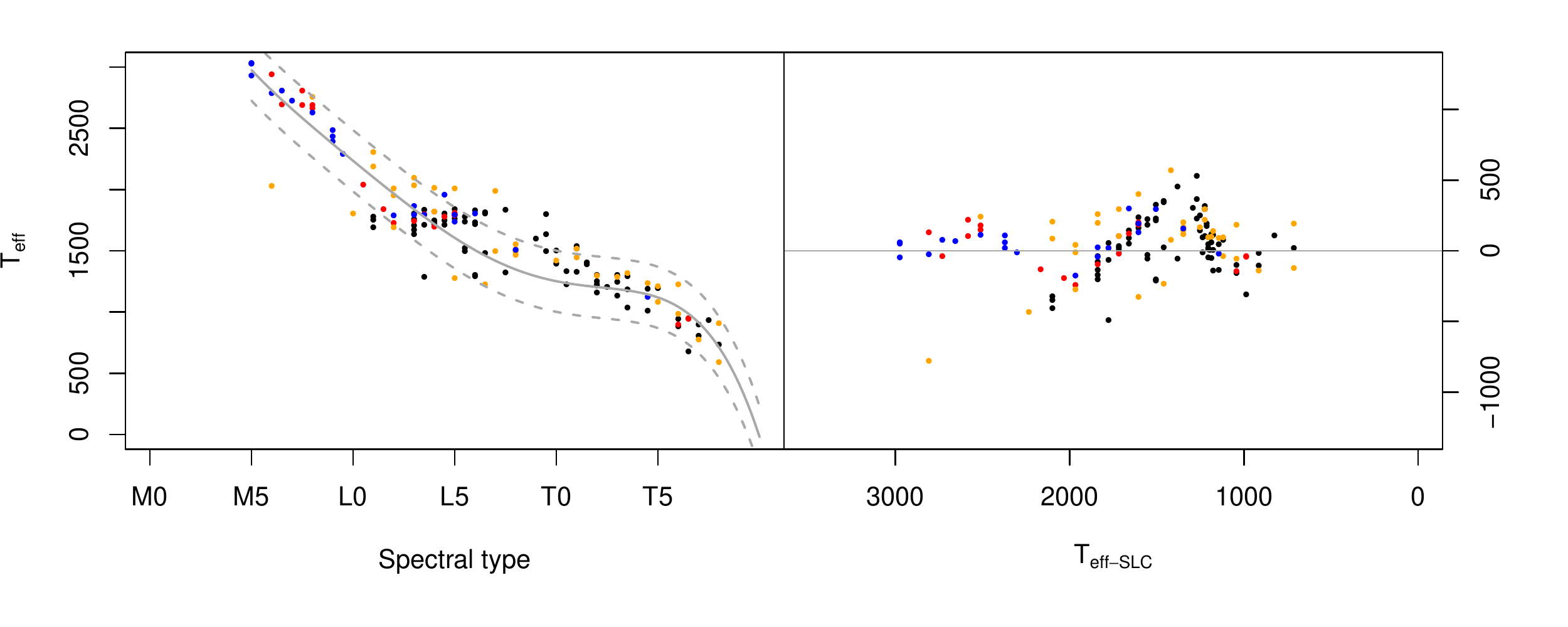}
  \caption{Effective temperatures derived from the best $\chi^2$ fits
    to BT-Settl models as a function of the spectral type assigned in
    the literature. Black circles correspond to the compilation by
    Leggett, red circles to the compilation of Keck LRIS spectra by
    Reid, orange circles to the NIRSPEC compilation, and blue ones to
    the IRTF compilation. The grey continuous line shows the
    {\teff}-spectral type calibration by \cite{2009ApJ...702..154S}
    from optical and infrared spectra. The dashed lines represent the
    same calibration $\pm$ 250 K. The right-hand panel shows the
    residuals ({\teff}$_{\rm (predicted)}$-{\teff}$_{(SLC)}$) with
    respect to the calibration.}
  \label{ji2fits}
\end{figure*}

The best $\chi^2$ fits to the full resolution spectra give us a lower
limit to the errors attainable in the {\Gaia} {\teff}
estimates. Figure \ref{ji2fits} shows in the $y$-axis the effective
temperatures derived from the best $\chi^2$ fits described in the
previous paragraph as a function of the spectral types assigned in the
libraries. We introduced a small jitter (characterised by a standard
deviation equal to 30K) in the values of the fitted temperature to
enhance the visibility of the stars with the same spectral type. In
the same plot we have included the calibration of effective
temperatures with spectral types by \cite{2009ApJ...702..154S} as a
continuous line. It corresponds to the calibration derived from
optical spectral types of L dwarfs and infrared spectral types of the
T dwarfs, valid in the M6 to T8 range (the SLC calibration). The SLC
calibration relies on model atmospheres described in
\cite{2008ApJ...689.1327S}, and references therein. The dashed lines
correspond to the same calibration plus/minus 250 K. The right-hand
panel shows the difference between the predicted temperature and the
temperature obtained applying the SLC calibration to the spectral
types assigned in the compilations, as a function of the latter. The
scatter of {\teff} values in Fig. \ref{ji2fits} around the SLC
calibration reflects errors of the regression process, but also the
uncertainty in the calibration between spectral types and effective
temperatures. In this work, we are taking the effective temperature
derived through the SLC relation as the ground truth. However, the SLC
relation, derived from high-resolution infrared spectra, which are much
more appropriate for determining an effective temperature than
the {\Gaia} spectrophotometry, has an intrinsic scatter estimated by
the authors as $\approx 100 K$ (although the article lacks details
about how the scatter was estimated, and visual inspection of the
plots suggests a much higher scatter at least in the L
sequence). Furthermore, \cite{2011ASPC..448..929R} compared a
calibration derived from the BT-Settl model library (the one used in
this work) with several other calibrations for the M spectral
subsequence in the literature, and found variations of the order of
200-300 K for each spectral subtype. \cite{2011A&A...536A..63B} showed
even higher values of the scatter for the same spectral type range
(they derived the {\teff} values from global fits to the SED). This
shows that the calibration of the relationship between effective
temperatures and spectral types is a problem that is not yet fully
solved, and that the exact {\teff} values derived for an UCD depend on
the calibration used, with typical uncertainties of a few hundred
Kelvin. The predictions of the regression modules presented in this
section have to be judged in this context. In this respect, our
regression models for the {\Gaia} data have the advantage that they
will provide UCD temperatures derived consistently from the BT-Settl
model family.

Three stars lacked precise spectral subtypes in the spectral
compilations, namely 2MASS 1237392+652615, SDSS1346464-003150, and
SDSS1624144+002916. We assumed them to be T6.5
\citep{2011ApJS..197...19K}, T6.5 \citep{2010A&A...522A.112R}, and T6
\citep{2011ApJS..197...19K}. Figure \ref{ji2fits} shows a
tendency to assign effective temperatures around 1700 K for sources
with spectral type L.

Table \ref{ji2disp} lists the mean difference $\mu$ between the
$\chi^2$ {\teff} estimates and the effective temperatures derived from
the SLC calibration and the spectral types cited in the spectral
compilations (hereafter bias); the standard deviation with respect to
the bias-corrected mean ($\sigma$) displayed by the four compilations;
and the RMSE without correcting for the mean bias.  We list these
values for the two model families (AMES Cond + Dusty and BT-Settl)
used in this work. All RMSE values are given in Kelvin.

\begin{table*}
\caption{Average bias ($\mu$) and standard deviation of the $\chi^2$
 effective temperature (K) fits. }
\label{ji2disp} 
\centering      
\begin{tabular}{c | c c c | c c c} 
\hline\hline             
Library & $\mu$ (BT-Settl) & $\sigma$ (BT-Settl) & RMSE (BT-Settl) & $\mu$ (AMES) & $\sigma$ (AMES) & RMSE (AMES)\\ 
\hline                       
Leggett & 42 & 196 & 199  &  156 & 197 & 250 \\ %
Reid    & 41 & 72  & 143  & -84 & 116 & 235 \\ %
NIRSPEC & 67 & 177 & 256  &  63 & 175 & 251 \\ %
IRTF    & 53 & 63  & 126  &  61 & 70 & 140 \\ %
\hline                       
\end{tabular}
\end{table*}

The compilations by Reid and Leggett have a good overall wavelength
coverage in the RP range, and spectra with poor coverage were removed
from the validation set. Since the expected emission in the wavelength
regions of the RP range not covered by the observed spectra (bluewards
of 750 nm) is negligible in the {\teff} and {\logg} parameter ranges
under consideration, we expect the completions to have little or no
effect on the subsequent parameter estimation with the models
described in the previous section. The NIRSPEC library is in the
opposite case, with a majority of spectra starting around 1.1
$\mu$m. At this wavelength, the RP transmission is close to zero, and
therefore, the input for the GOG simulations of the {\Gaia} RP spectra
in the wavelength region where the transmission is high comes only
from a synthetic model (the best $\chi^2$ match) that was actually
used during the training phase of the algorithms. As a consequence,
the errors in the parameter estimates for these stars are overly
optimistic. The IRTF library is an intermediate case with most spectra
covering the wavelength region above 0.8 $\mu$m where most of the
source flux is concentrated. Thus, the completion will not have a
relevant impact on the resulting simulated RP spectrum.

\begin{figure*}[thb]
  \centering
  \includegraphics[scale=1]{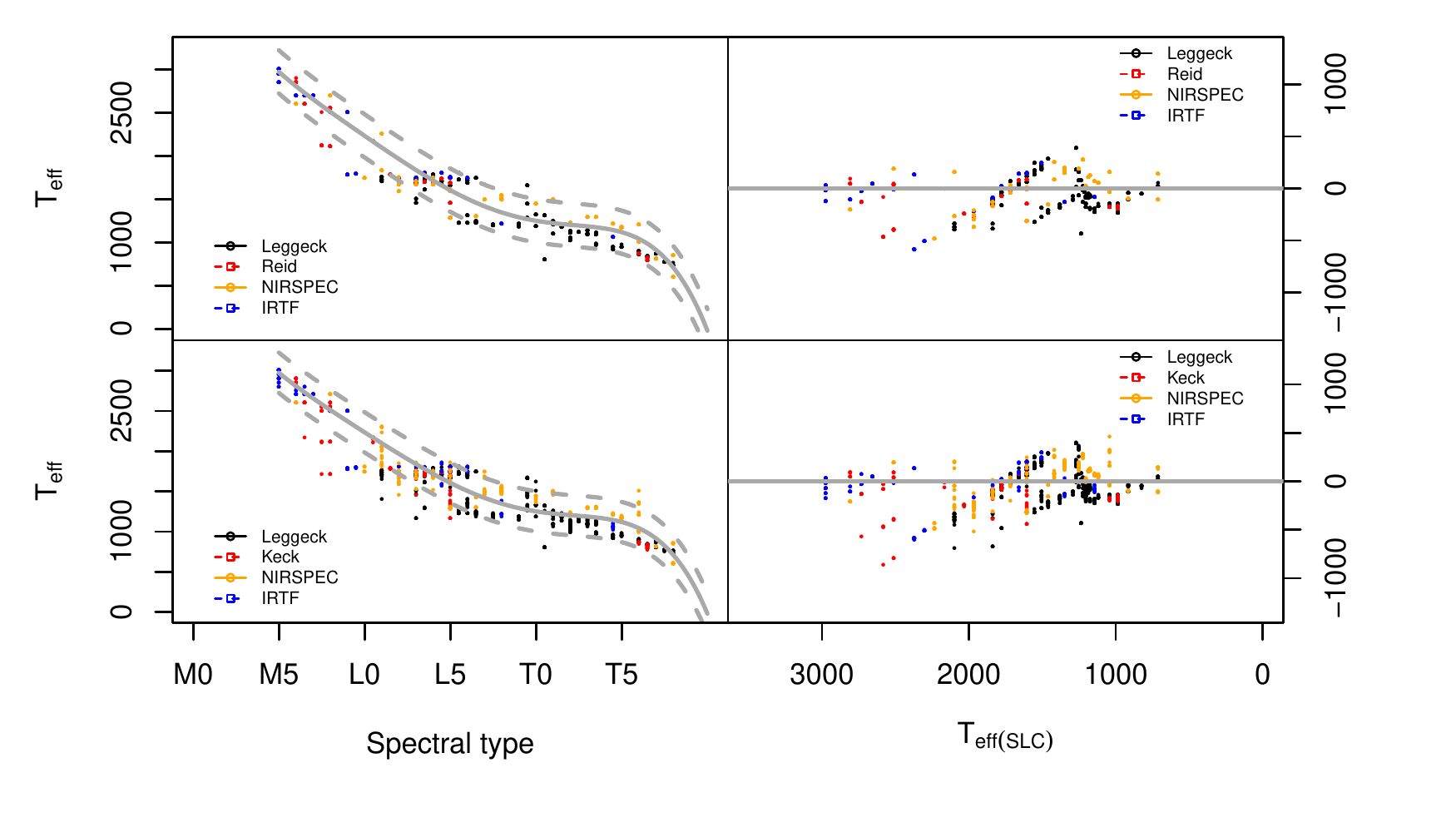}
  \caption{{\teff} predictions obtained by the kNN
    algorithm based on the BT-Settl grid of models for the four
    libraries of ground-based spectra listed in section \ref{valid}
    for G=15 (top row) and G=20 (bottom row). The colour code is the
    same as used in Fig. \ref{ji2fits}. The $x$-axis shows the spectral
    types gathered from the literature cited in the spectral
    compilations. The grey line shows the {\teff}-spectral type
    calibration by \cite{2009ApJ...702..154S} from optical and
    infrarred spectra ($\pm250$K, grey dashed lines). The right-hand
    panels show the residuals with respect to the calibration as in
    Fig. \ref{ji2fits}.}
  \label{pcaknng20}
\end{figure*}

\begin{figure*}[thb]
  \centering
  \includegraphics[scale=1]{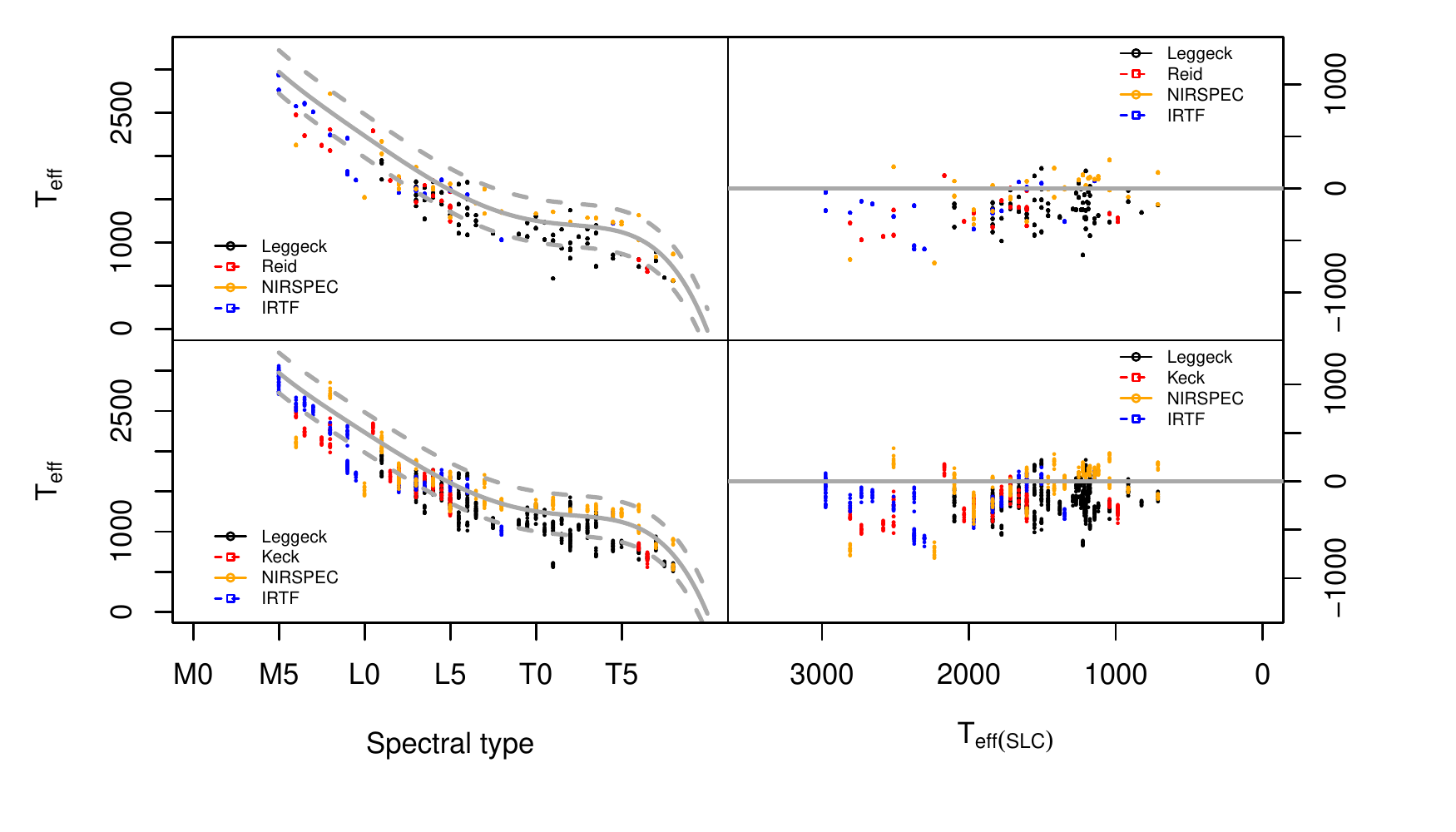}
  \caption{{\teff} predictions obtained by the GP model
    based on the BT-Settl grid of models for the four libraries of
    ground-based spectra listed in section \ref{valid} for G=15 (top
    row) and G=20 (bottom row). The colour code is the same as used in
    Fig. \ref{ji2fits}. The $x$-axis shows the spectral types gathered
    from the literature cited in the spectral compilations. The grey
    line shows the {\teff}-spectral type calibration by
    \cite{2009ApJ...702..154S} from optical and infrared spectra
    ($\pm250$K, grey dashed lines). The right-hand panels show the
    residuals with respect to the calibration as in
      Fig. \ref{ji2fits}.}
  \label{gpg20b}
\end{figure*}

Figures \ref{pcaknng20} and \ref{gpg20b} show the predictions of the
kNN and GP models for the empirical libraries described
in previous paragraphs. The predicted values of {\teff} are compared
with the effective temperature assigned by the SLC calibration to the
spectral type provided by the empirical libraries. The model used in
all panels corresponds to the one trained with noisy spectra of G=20
and 70 transits. The upper panels describe the performance of the
model when applied to noisy spectra corresponding to G=15 and 70
transits whilst the lower panel shows the results for the same model
applied to GOG simulations at G=20 and 70 transits. For each spectrum
simulated by GOG, we constructed ten noisy replicates using the Gaia
error model currently available, and predict the values of {\teff} and
{\logg} for each one of them. The scatter in the predictions for these
noisy replicates is better visible in the lower panels (spectra
simulated at G=20).

\begin{table*}[thb]
\caption{Root mean square errors (RMSE) of the GP
  and kNN models for the prediction of {\teff}, applied to the
  compilations of empirical spectra.}
\label{GPKNNRMSE} 
\centering      
\begin{tabular}{l | c c c | c c c} 
\hline\hline             
Number of transits &  \multicolumn{3}{c|}{28} & \multicolumn{3}{c}{70}\\
\hline                       
G & 15 & 18 & 20 & 15 & 18 & 20 \\
\hline                       
GP-G20         &257 (192)&260 (194)&266 (201)&256 (191)&257 (192)&260 (196)\\
PCA-GP-G20     &266 (199)&270 (203)&281 (216)&264 (197)&266 (200)&272 (207)\\
kNN-G20        &209      &210      &213      &207      &209      &213 \\
PCA-kNN-G20    &211      &213      &215      &209      &207      &210 \\
Bayes BT-Settl       & 230.5  & 235.7  & 239.0  & 243.4  & 241.6  & 239.7 \\
Bayes COND/DUSTY     & 252.6  & 252.3  & 255.0  & 257.5  & 257.6   & 258.0 \\
\hline\hline                        
\end{tabular}
\end{table*}

Table \ref{GPKNNRMSE} collects the root mean squared errors (RMSE) of
the {\teff} estimates (expressed in Kelvin) obtained by the GP and kNN
models trained with a collection of synthetic spectra of the BT-Settl
library simulated for an apparent magnitude G=20. The models are then
applied to the four libraries of empirical spectra simulated at 28 and
70 transits, and for apparent magnitudes G=15, 18, and 20. The prefix
PCA refers to the models built for the input space of principal
componets. The values in parenthesis correspond to the RMSE after
correcting for the systematic bias in GP predictions. We also include
the RMSE of the Bayesian inference modules built from the BT-Settl and
COND/DUSTY model libraries for comparison.

\subsection{Bayesian inference} 
\label{sub:estimates_from_sampling}

\begin{figure*}[thb]
  \centering
  \includegraphics[scale=1]{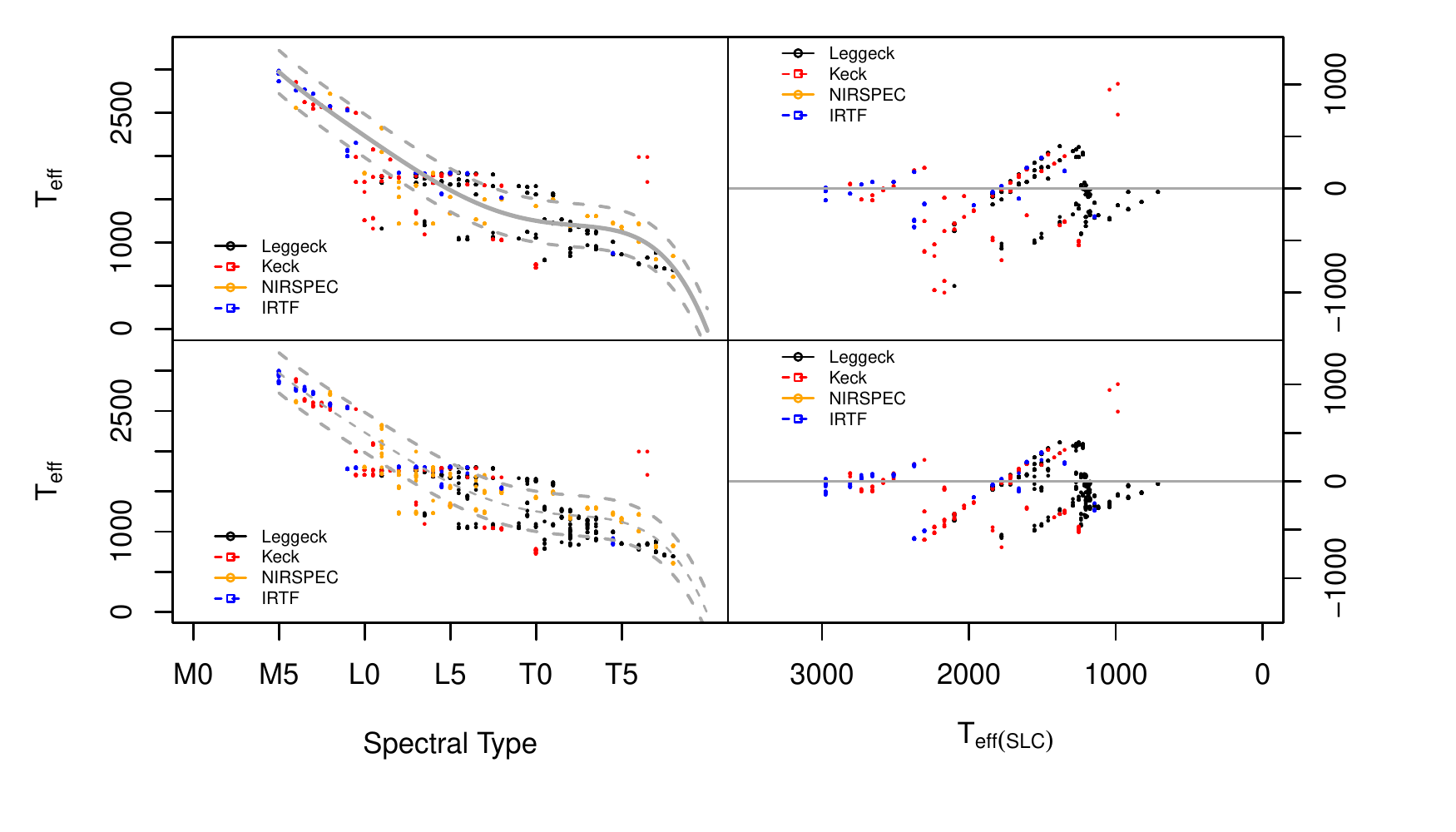}
  \caption{{\teff} predictions obtained from the ellipsoidal samplings
    (BT-Settl) for the four empirical libraries of ground-based
    spectra. The colour code is the same as used in Fig.
    \ref{ji2fits}. The $x$-axis shows the spectral types gathered from
    the literature cited in the spectral compilations. The grey line
    shows the {\teff}-spectral type calibration by
    \cite{2009ApJ...702..154S} from optical and infrared spectra
    ($\pm250$K, grey dashed lines). The right-hand panels show the
    residuals with respect to the calibration as in
      Fig. \ref{ji2fits}. The top row corresponds to the spectra
    simulated at G=15 and the bottom row corresponds to the G=20
    replicates.}
  \label{fig:bayes}
\end{figure*}

Figure~\ref{fig:bayes} shows the {\teff} predictions obtained from the
ellipsoidal samplings, using the neural network trained with the
BT-Settl grid of models and the current error model for the {\Gaia} RP
spectra. The predictions are obtained as before, for ten replicates of
each GOG simulation of an empirical spectrum. It shows a tendency to
predict effective temperatures around 1700 K for stars with spectral
type L. Furthermore, there are indications of a second attractor
slightly above 1000 K for stars and brown dwarfs with spectral types
between L5 and T5. These systematic biases can also be recognised in
Figs.  \ref{ji2fits} (representing the {\teff} estimates from $\chi^2$
fits) and \ref{pcaknng20} (representing the {\teff} estimates from $k$-nearest neighbours), but not in Fig. \ref{gpg20b} which corresponds to
the GP model. The fact that the Bayesian inference shows
the same kind of biases as nearest neighbours and the $\chi^2$ fits is
not surprising because Bayesian inference is equivalent to maximum
likelihood estimation under flat priors such as those used in this
work. Minimum $\chi^2$ fits and nearest neighbours are special cases
of maximum likelihood estimation.

\begin{figure*}[thb]
  \centering
  \includegraphics[scale=.65]{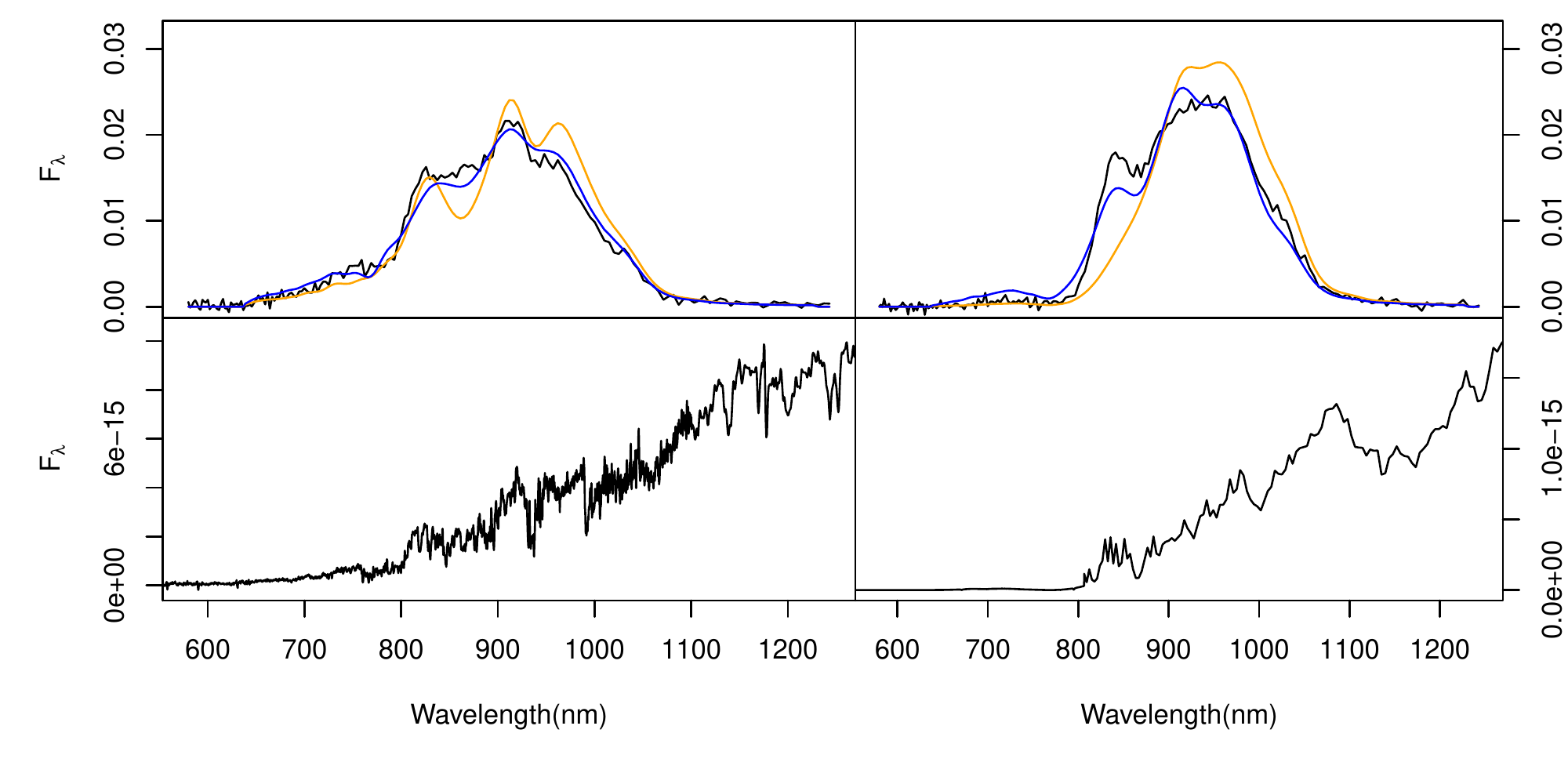}
  \caption{(Top panel, black continuous lines) GOG simulations of
    spectra contained in the Legget compilation corresponding to
    2MASS0345+25 (left) and SDSS1511+06 (right) for G=20 and 70
    transits. The blue line represents in both panels the model
    corresponding to the mode of the posterior probability as derived
    using ellipsoidal sampling. In orange, the model that corresponds
    to the effective temperature derived from the spectral type (L1
    and T0 respectively) using the SLC calibration. The lower row
    shows the original spectra completed with the best $\chi^2$
    model.}
  \label{fig:1700}
\end{figure*}

To understand the nature of this trend to concentrate the
predictions by the Bayes module around 1700-1800 K, we plot in Fig.
\ref{fig:1700} the best fits produced by the Bayesian inference
module (blue), the model corresponding to the spectral type assigned
in the spectral compilations (orange), and finally, one of the ten
noisy replicates of the spectrum simulated with GOG for G=20 (black),
for two stars with spectral types L1 (left) and T0 (right), both of
which are predicted to have temperatures between 1600 and 1700 K. For
these two stars we also plot the original spectra for reference
(bottom row).

\begin{figure*}[thb]
  \centering
  \includegraphics[scale=0.65]{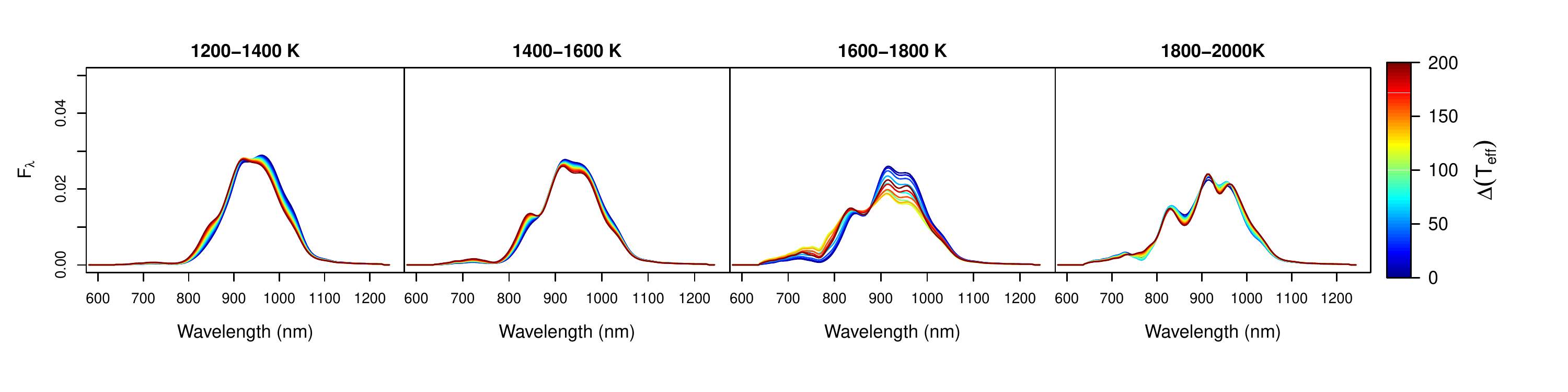}
  \caption{BT-Settl model spectra between 1200 and 2000 K (the range
    of temperatures where the Bayesian module shows a clear tendency
    to concentrate predictions around 1600-1800 K) and
    {\logg=5.0}. Each panel shows spectra in an interval of
    temperatures of 200 K. The colour code represents the increment in
    effective temperature with respect to the lowest temperature
    covered in the panel. Blue lines correspond to this lowest
    temperature and red continuous lines correspond to the lowest
    temperature plus 200 K. }
  \label{fig:plateaux}
\end{figure*}

The reason for this systematic effect seems related to the fact that
between 1600 and 1800 K, the model spectra undergo, especially at
{\logg}=5.0, rapid changes with temperature, whilst before and after
this range, we find plateaux where the spectra have only a mild
dependence with temperature (see Fig. \ref{fig:plateaux}). From the
sampling perspective, relatively strong changes in the proposed
temperatures falling within the plateaux regions result in small
changes in the likelihood. Sampling in the 1600-1800 K region on the
contrary provides a more varied range of models susceptible to better fit
the observed noisy spectrum.

\begin{figure*}[thb]
  \centering
  \subfigure[]{\label{fig:41}
    \includegraphics[scale=0.33]{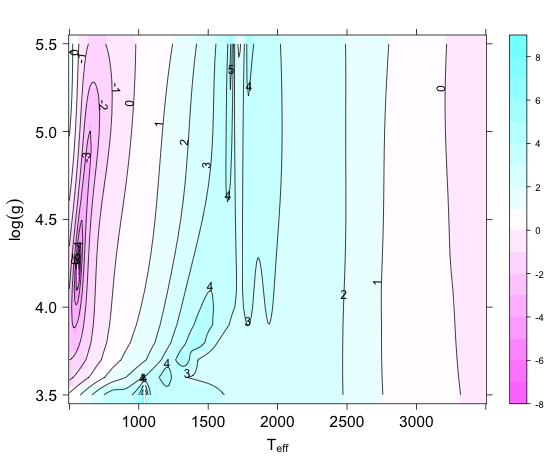}}
  \subfigure[]{\label{fig:64}
    \includegraphics[scale=0.33]{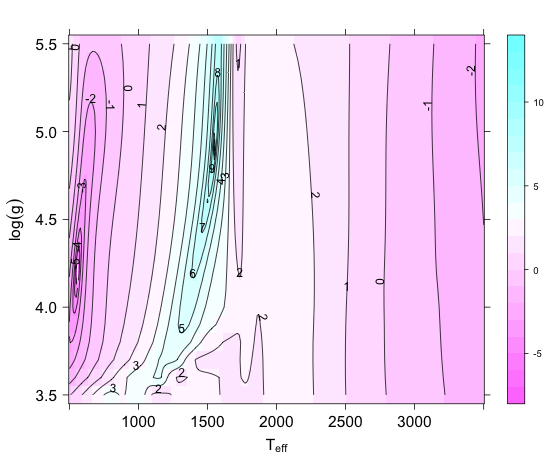}}
  \caption{Log-likelihood landscapes for a G=20 noisy replicate of the
    spectrum of SDSS0107 (left), and the BT-Settl model for 1550 K and
    {\logg}=5.0 (right).}
  \label{fig:likelihoods}
\end{figure*}

The detailed analysis of the predictions reveals a large variance
within some of the blocks of ten noisy replicates of a given
spectrum. A good example of this kind of problems is provided by SDSS
J010752.33+004156.1, where six estimates cluster around 1036 K whilst
the remaining four cluster around 1664 K. As a reference, the spectral
type quoted for this object is L5.5 or, equivalently, 1554 K according
to the SLC calibration. This apparent inconsistency is due to the
multimodal posterior distribution, which is closely related to the
likelihood landscape for flat priors and noisy
spectra. Figure~\ref{fig:41} shows the log-likelihood distribution
(derived from one noisy GOG simulation of the spectrum in the Leggett
compilation) as a function of the parameters for SDSS
J010752.33+004156.1. For comparison, we show the equivalent plot for
the model in the BT-Settl grid corresponding to $\bm{\theta} = (1550
K, 5.0)$ (Fig. \ref{fig:64}). Whilst the log-likelihood landscape for
the BT-Settl model is unimodal, the log-likelihood landscape for the noisy spectrum of SDSS J010752.33+004156.1 shows maxima of
comparable height at 1036 and 1664 K (in addition to other local
maxima in the range 1000 K $<$ {\teff} $<$ 1500 K, 3.5 $<$ {\logg} $<$
4.0). Depending on the particular realisation of the noise, the
ellipsoidal sampling will converge to different maxima. In practice,
we find cases of clear bimodality in spectra of stars with spectral
types between L1 and T2 ({\teff} between 1200 and 2000 K).

\begin{figure*}[thb]
  \centering
    \includegraphics[scale=0.4]{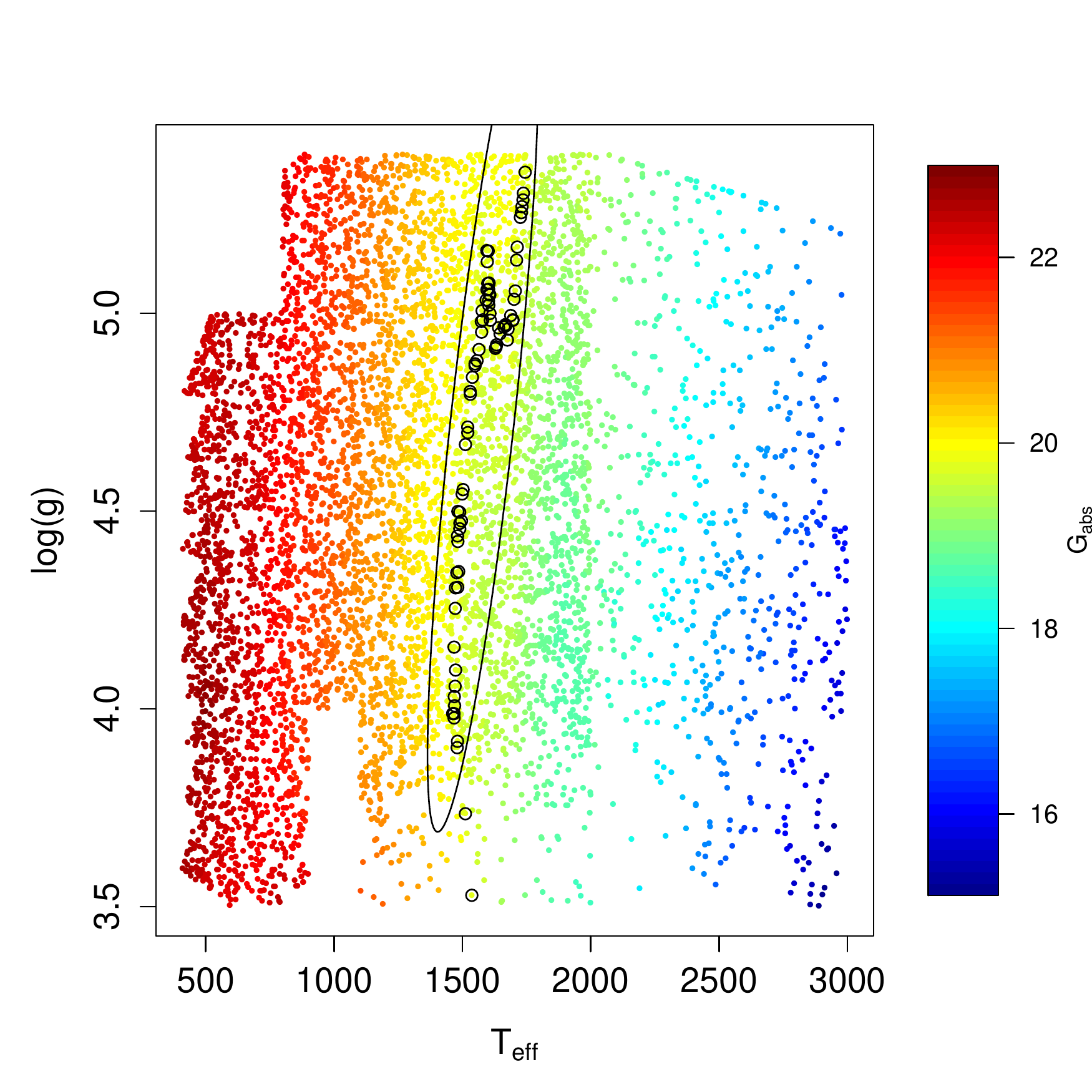}
  \caption{Absolute G magnitudes for the BT-Settl model library as a
    function of the physical parameters {\teff} and {\logg} (see
    colour code at the right-hand side of the scatter plot). Models
    within $G_{\rm abs}=19.8492_{-0.0194}^{+0.0191}$ (corresponding to
    SDSS J010752.33+004156.1 simulated at G=20.0) are shown as black
    circles. The ellipse shows the 1-$\Sigma$ iso-contour of the
    Gaussian physical prior described in the text.}
  \label{prior}
\end{figure*}

This problem is easily solved by using priors based on the additional
information provided by the {\Gaia} astrometric measurements, as
suggested in Sect. \ref{method}. For this case, Fig. \ref{prior} shows
a particular choice of the prior that would discard all local maxima that are
inconsistent with the {\Gaia} photometry/astrometry, under the
assumption of negligible circumstellar and interstellar extinction. In
it, we represent the distribution of absolute G magnitudes of the
BT-Settl models as a function of the dependent parameters {\teff} and
{\logg}. The continuous line shows the 1-$\Sigma$ countour of the
two-dimensional Gaussian prior (with $\Sigma$ being the covariance
matrix). The prior is fully defined by the mean $\mu_{prior}$ and
covariance $\Sigma_{prior}$ of the Gaussian prior distribution. Given
a set of astrometric, photometric and spectrophotometric observations
such as those simulated for G=20.0 with GOG for SDSS
J010752.33+004156.1, and the photometric and astrometric errors shown
in Fig. \ref{sigmaGpi}, we calculate an absolute G magnitude equal to
$19.8492_{-0.0194}^{+0.0191}$, where the uncertainties are derived
from the values shown in Fig. \ref{sigmaGpi} multiplied by five. We
used five times the nominal uncertainties in the apparent G magnitude
and $\pi$ (the parallax) to account for the potential
mismatch of the model-predicted magnitudes with respect to the real
distribution. We did not take the Lutz-Kelker bias into account in
the computation of the uncertainties of the absolute G
magnitude. Models within the uncertainties in the absolute G magnitude
($19.8492_{-0.0194}^{+0.0191}$) are shown in Fig. \ref{prior} as black
circles. We used the {\teff} and {\logg} parameters of these models to
propose the values of $\mu_{prior}$ and $\Sigma_{prior}$ that were
used to draw the 1-$\Sigma $ iso-countour in Fig. \ref{prior}. This
prior is eleven orders of magnitude larger at the 1664 K maximum than
at the 1036 K one, and thus renders this local maximum insignificant
in the posterior probability density distribution. There is a narrow
local maximum at {\teff}$\approx$1500 K and {\logg}$\approx$4.9 in
Fig. \ref{fig:likelihoods} where the prior is only one order of
magnitude smaller than that corresponding to the 1664 K, but this
maximum is never significantly sampled in any of our ten replicates
due to its narrowness.

\subsection{Model selection}

Ellipsoidal sampling allows estimating the evidence given
an observed spectrum and a model choice (COND+DUSTY or BT-Settl), as
defined in Eq. \ref{eq:evidence1}. Therefore, it is possible to
compare these two model families from a Bayesian perspective. \(
M_0 \) denotes the BT-Settl model library and $M_1$ the COND+DUSTY
library. If these were the only two alternatives, then the Bayes
factor (BF) in support of the $M_0$ model would be defined as the
ratio of the respective marginal densities (evidences) of the data for
the two models,

\begin{equation}
BF = \frac{m_0(\mathbf{s})}{m_1(\mathbf{s})}.
\end{equation}

If \( \pi_0 \) and \( \pi_1 \) denote the respective prior
probabilities (in our case \( \pi_0 = \pi_1 = 1/2 \)), the
posterior probability of \( M_0 \) is given by

\begin{equation}\label{eq:comparingmodels}
  \Pr (M_0|\mathbf{s}) = \frac{\pi_0 BF}{\pi_0 BF + \pi_1}.
\end{equation}

We obtain that 73.4~\% of the spectra support the BT-Settl model
library against 26.6~\% supporting the COND-DUSTY combination. For
this reason we have exemplified the regression results with figures
and discussion related to the models obtained with the BT-Settl
library.

\subsection{Estimates of the gravity}

\begin{figure*}[thb]
  \centering
  \includegraphics[scale=.75]{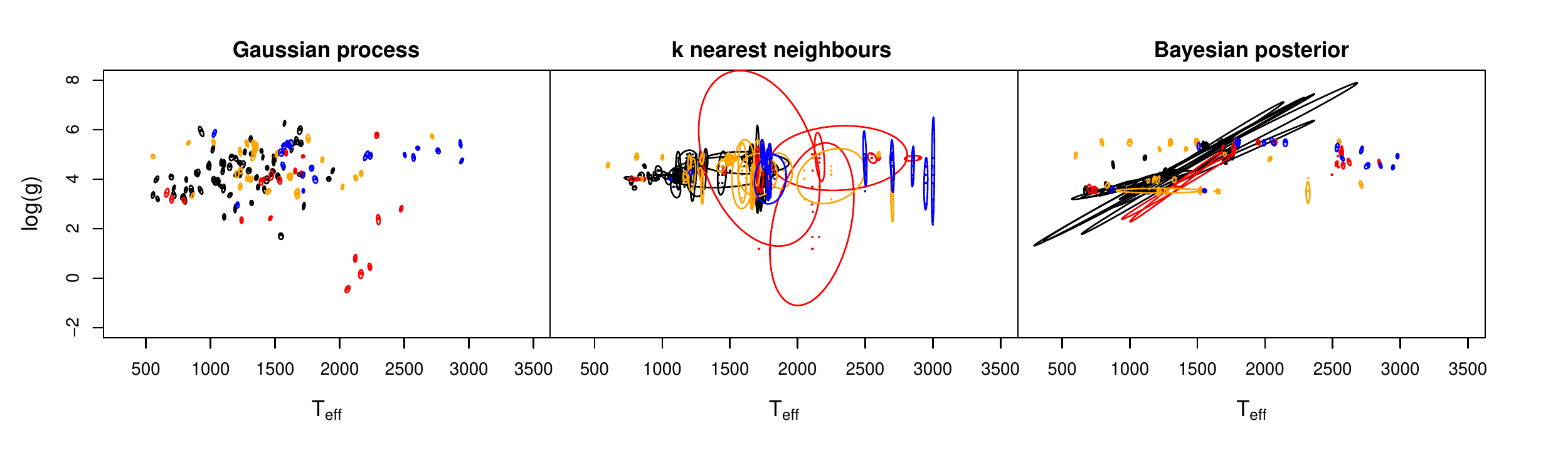}
  \caption{{\teff}--{\logg} predictions for the empirical spectral
    libraries obtained with the GP model (left), the
    kNN model (middle), and the Bayesian inference
    (right). The colour code is the same as in Fig. \ref{ji2fits}. The
    ellipses correspond to the covariance estimated from ten noisy
    replicates of the GOG simulated spectra (G=15).}
  \label{fig:loggDM}
\end{figure*}

\begin{figure*}[thb]
  \centering
  \includegraphics[scale=.75]{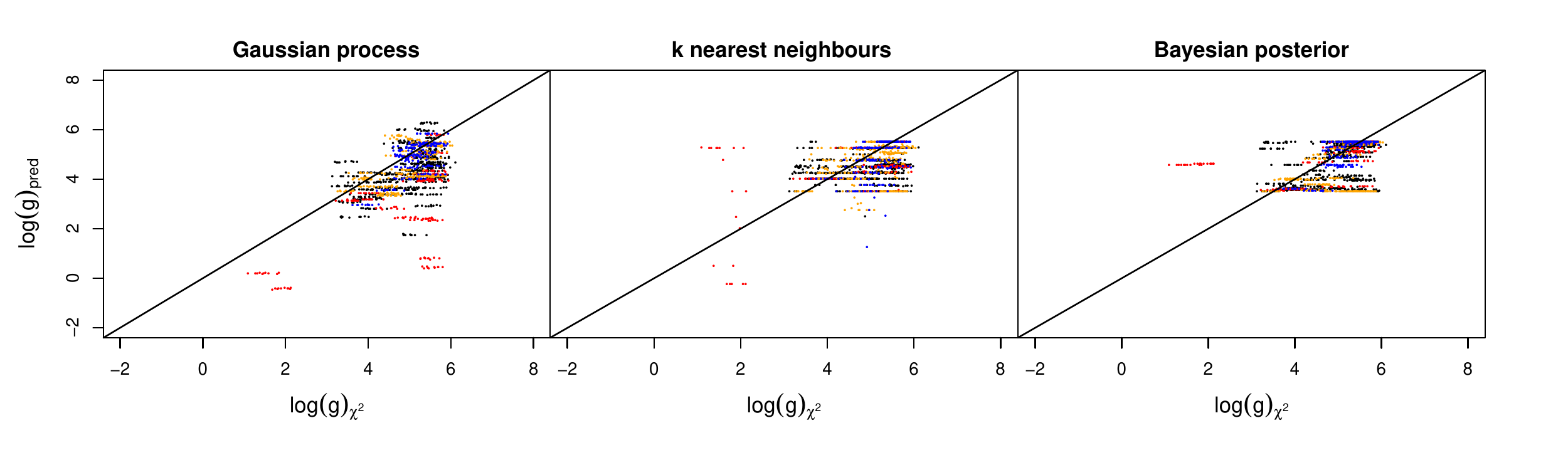}
  \caption{{\logg} predictions for the empirical spectral libraries
    obtained with the GP model (left), the kNN model (middle), and the Bayesian inference (right), as a
    function of the {\logg} value of the minimum $\chi^2$ fit
    (jittered with a Gaussian distribution of $\sigma=0.2$). The
    colour code is the same as in Fig. \ref{ji2fits}. }
  \label{fig:loggDM2}
\end{figure*}

Figure \ref{fig:loggDM} shows the {\teff}--{\logg} diagrams obtained
from the three models discussed in this section. The ellipses
represent the covariance estimated from the ten noisy replicates at
G=15 of each GOG simulations of the empirical spectra. Since we do not
have a compilation of surface gravities available to assess the
overall validity of the predictions, we compare the regression values
with those obtained from the $\chi^2$ fitting to the full resolution
spectra. Figure \ref{fig:loggDM2} shows this comparison. The $\chi^2$
fit values are jittered with a Gaussian distribution of standard
deviation equal to 0.2 to enhance the visibility. If the
$\chi^2$ estimates are taken as targets, only the GP
model can be used to obtain very rough estimates of the gravity and to
tag low-gravity candidates.

The only star in our samples with an indication of low gravity in the
comments section of the Dwarf
Archives\footnote{\url{http://spider.ipac.caltech.edu/staff/davy/ARCHIVE/index.shtml}}
is the NIRSPEC target star 2MASS J1726000+153819. For this star, both
the kNN model and the Bayesian estimate agree to assign a value of
{\logg}=3.5, while the GP model assigns a higher value {\logg}=4.1.

\section{Conclusions\label{conclusions}}

We have presented the module that will be in charge of detecting and
characterising ultra-cool dwarfs in the Gaia database. The module is
subject to change and improvement, but this implementation provides
the baseline performance that can be expected from it.

We used the current instrument models and the estimated spatial
densities by \cite{2008A&A...488..181C} to predict the expected number
of ultra-cool dwarfs per spectral type bin. We found that {\Gaia} will
be able to detect significant numbers (around or above ten detections)
for UCDs of spectral types hotter than L6-7 {\sc v}. We also used
the BT-Settl library of synthetic spectra to define selection
criteria for the UCD module such that no UCD is missed due to
measurement errors. Given the spatial densities estimated by
\cite{2008A&A...488..181C}, we derived contamination rates from stars
hotter than 2500 K in the resulting samples.

We conducted an extensive study to find the best
statistical regression model of the relationship between the observed
{\Gaia} RP spectrum and the source physical parameters ({\teff} and
{\logg}). We evaluated several alternatives in view of their
internal and external errors. The internal errors were estimated with
cross validation experiments with a dataset interpolated from the
nominal grid of models provided by the libraries of synthetic
spectra. The external validation was carried out by applying the
regression models to an independent set of UCD spectra observed from the
ground. All these experiments were carried out on GOG simulations of
the full-resolution spectra for a number of apparent G magnitudes and
numbers of transits. 

As a result, we found that the expected end-of-mission error of the UCD
module for the faintest detectable UCDs (G=20) is 210 K for the kNN
module and 260 for the GP module (207 K if a bias correction is
applied). These performances are approximately constant as a function
of G, at least down to G=15 (a typical apparent magnitude for the
  brightest UCDs in the Gaia catalogue), and are remarkably close to the
performance of a simple $\chi^2$ fit with the full-resolution
spectrum.

The Bayesian inference of the source parameters shows systematic
deviations in the distribution of predicted temperatures, which are
also apparent (although less conspicuous) in the $\chi^2$ fits and kNN
predictions. It is also severely affected by the multimodality of the
likelihood maps. The application of physical priors and advanced
sampling techniques capable of identifying multiple modes in the
posterior will be the subject of a forthcoming paper in this series.

The logg predictions are characterised by a typical RMSE of 0.2 dex
for the GP module (0.7 dex for the kNN module) as measured by the
cross-validation experiments. Unfortunately, these error estimates are
overly optimistic because they are derived from testing the
regression modules on synthetic spectra and not on observed spectra of
real UCDs. The {\logg} estimates for the empirical spectra observed
from the ground, although broadly consistent with the expected
distribution of values in the samples of empirical spectra, prove that
the errors quoted above are indeed extremely optimistic. The GP module will undoubtedly benefit from a more realistic
(i.e. wavelength-dependent) treatment of the noise parameter.


\begin{acknowledgements}

The authors wish to acknowledge the Coordination Unit 2 of the Gaia
DPAC for the use of the GOG simulator, and Rosanna Sordo for their
kind help and guidance with the simulation of both synthetic and
empirical spectra at the {\Gaia} instrumental characteristics. we would also
 like to thank the referee, Coryn Bailer-Jones, for the
insightful comments that significantly improved the first version of
the manuscript. LS aknowledges Jos{\'e} Caballero for his suggestions
regarding the estimation of the number counts of detectable UCDs
according to his volume density estimation. This research has been
supported by the Spanish Ministry of Science through grants
AyA2011-24052, AyA2010-21161-C02-02, AyA2009-14648-C02-01, CONSOLIDER
CSD2006-00070, CSD2007-00050, and PRICIT-S2009/ESP-1496.

\end{acknowledgements}

\bibliographystyle{aa} \bibliography{sarro}

\end{document}